%% file: paper.tex
\documentclass[letterpaper,twocolumn,10pt]{article}
\usepackage{usenix2019_v3}

\usepackage{tikz}
\usepackage{listings}
\usepackage{subcaption}
\usetikzlibrary{automata,arrows,positioning,snakes,decorations.pathreplacing,fit,tikzmark,decorations.text}

\usepackage{amsmath}

\include{header}

\usepackage{titling}
\iffalse
  \usepackage[compact]{titlesec}
  \usepackage{enumitem}
  \setlist{nolistsep}
\else
  \usepackage[moderate]{savetrees}
\fi

\usepackage{draftwatermark}
\SetWatermarkText{DRAFT---Please do not distribute}
\SetWatermarkScale{.15}
\SetWatermarkAngle{0}
\SetWatermarkText{}
\SetWatermarkVerCenter{30pt}

\newcommand{\syscall}{\Code{syscall}\xspace}
\newcommand{\syscalls}{\Code{syscall}s\xspace}
\newcommand{\stramp}{nexpoline\xspace}
\newcommand{\srgadget}{\emph{sysret-gadget}\xspace}
\newcommand{\sysretg}{\Code{syscall; return;}\xspace}

\newcommand{\fork}{\Code{fork}\xspace}
\newcommand{\exec}{\Code{exec}\xspace}
\newcommand{\sudo}{\Code{sudo}\xspace}

\newcommand{\domain}{domain\xspace}
\newcommand{\domains}{domains\xspace}
\newcommand{\segment}{subspace\xspace}
\newcommand{\segments}{subspaces\xspace}
\newcommand{\subdomain}{endoprocess\xspace}
\newcommand{\Subdomain}{Endoprocess\xspace}
\newcommand{\subdomains}{endoprocesses\xspace}

\newcommand{\srand}{secc\_rand\xspace}
\newcommand{\semph}{secc\_eph\xspace}
\newcommand{\sscet}{secc\_cet\xspace}
\newcommand{\sdcet}{disp\_cet\xspace}
\newcommand{\sdemph}{disp\_eph\xspace}

\newcommand{\pkru}{\Code{PKRU}\xspace}
\newcommand{\wrpkru}{\Code{WRPKRU}\xspace}
\newcommand{\ud}{untrusted \domain}
\newcommand{\td}{trusted \domain}
\newcommand{\xcall}{\Code{xcall}\xspace}

\newcommand{\ptrace}{ptrace\xspace}

\newcommand{\sphere}{protection sphere\xspace}
\newcommand{\Sphere}{Protection Sphere\xspace}
\newcommand{\xswitch}{domain-switch\xspace}

\newtheorem{prop}{Property}
\newtheorem{defi}{Definition}

\newcommand{\visor}{Endokernel\xspace}
\newcommand{\arch}{\visor Architecture\xspace}
\newcommand{\system}{Intravirt\xspace}

\newcommand{\nexpoline}{nexpoline\xspace}
\newcommand{\Nexpoline}{Nexpoline\xspace}

\newcommand{\libos}{libsep\xspace}

\newcommand{\monitor}{endokernel\xspace}
\newcommand{\Monitor}{Endokernel\xspace}

\newcommand{\nemo}{nested \monitor}
\newcommand{\nemos}{\nemo{}s\xspace}
\newcommand{\NEMO}{Nested \Monitor}
\newcommand{\NEMOS}{\NEMO{}s\xspace}

\newcommand{\model}{\Subdomain Model\xspace}
\newcommand{\boxing}{nested boxing\xspace}
\newcommand{\Boxing}{Nested boxing\xspace}
\newcommand{\boxes}{nested boxes\xspace}
\newcommand{\sandbox}{sandbox\xspace}

\newcommand{\unbox}{unbox\xspace}

\newcommand{\safebox}{safebox\xspace}

\newcommand{\grant}{\emph{grant}\xspace}
\newcommand{\revoke}{\emph{revoke}\xspace}

\tikzset{%
roundnode/.style={circle, draw=black, fill=gray!5, thick, minimum size=10mm},
roundnodeL/.style={circle, draw=black, fill=gray!5, thick, minimum size=15mm},
}

\title{\Large \bf \system: System Call and Signal Virtualization for
Secure Intra-Process Isolation \\on Leaky Operating System
Abstractions}
\title{\Large \bf \system: System Call and Signal
Virtualization for Secure and Efficient Nested Execution
Environment}
\title{\Large \bf \system: Syscall and Signal Virtualization for
Secure Intra-Process Isolation}
\title{\Large \bf \system: Securing Intra-Process Isolation
with Syscall and Signal Virtualization}

\title{\Large \bf \system: Intra-Process Privilege
  Separation for Syscall and Signal Virtualization}

\title{\Large \bf \system: A Nested Execution Environment
  with Syscall and Signal Virtualization}

\title{\Large \bf \system: A User-Level Privilege Separation
Framework for Memory, Syscall, and Signal Virtualization}

\title{\Large \bf \system: Nested Privilege
  Separation with Syscall and Signal Virtualization}

\title{\Large \bf \system: A Nested Execution Environment
  for Intra-Process Privilege Separation with Syscall and
Signal Virtualization}

\title{\Large \bf \system: %A Nested Separation Kernel for
Intra-Process Privilege Separation with Syscall and Signal
Virtualization}

\title{\Large \bf \system: Virtualizing System Interfaces
for Secure Intra-Process Privilege Separation}

\title{\Large \bf \system: Nested Security Monitor for
Syscall and Signal Virtualization}

\title{\Large \bf \system: Nesting Syscall and Signal
Virtualization for Secure Intra-Process Isolation}

\title{\Large \bf \system: A Nested Intra-Process Monitor
for Memory, Syscall, and Signal Virtualization}

\title{\Large \bf \system: A Nested Intra-Process Isolation Framework/Monitor
for Memory, Syscall, and Signal Virtualization}

\title{\Large \bf \system: Abstracting and Enforcing
Intra-Process Isolation Across Leaky Operating System
Boundaries}

\title{\Large \bf \system: Least-Authority Across Leaky Intra-Process
Isolation Abstractions}

\title{\Large \bf Programmable, Least-Authority Across Leaky \NEMO Abstractions}
\title{\Large \bf Preventing Leaky Abstractions in \NEMOS}
\title{\Large \bf Secure and Programmable \NEMOS}
\title{\Large \bf Secure and Programmable Nested Intra-Process Isolation}
\title{\Large \bf Secure and Programmable Nested Privilege Separation}
\title{\Large \bf Securing Leaky Abstractions and Enhancing Programmability for
\\Nested Privilege Separation}

\title{\Large \bf Enhancing the Safety and Programmability of Nested Privilege
Separation}

\title{\Large \bf Nested Privilege Separation}

\title{\Large \bf The \arch: An Extensible Kernel Architecture for Efficient and
Programmable Intra-Process Isolation}

\title{\Large \bf The \visor: A Nested Kernel Process Architecture for
  Efficient and Extensible Intra-Process Isolation}

\title{\Large \bf The \visor: A Nested Process Model for
  Efficient and Extensible Intra-Process Isolation}

\title{\Large \bf The \visor: A Process Model for Efficient and
Extensible Intra-Process Isolation}

\title{\Large \bf The \visor: An Operating System Architecture for Fast and
Extensible Single-Address Space Isolation}

\title{\Large \bf The \visor: A Nested Kernel Architecture for Fast and
Extensible Multidomain Subprocess Isolation}

\title{\Large \bf The \visor: A Nested Kernel Architecture for Fast and
Extensible Intra-Process Isolation}

\title{\Large \bf The \visor: A Kernelized Process Architecture for Fast
and Extensible Intra-Process Isolation}

\title{\Large \bf The \visor: A Kernelized Process Architecture for Fast and
Extensible Subprocess Virtualization}

\title{\Large \bf The \visor: A Kernelized Process Architecture for Fast and
Extensible Subprocess Virtual Machines}

\title{\Large \bf The \visor: A Process Architecture for Fast, Secure, and
Extensible Nested Virtualization}

\title{\Large \bf The \visor: Fast, Secure, and Extensible Process Virtualization}

\title{\Large \bf The \visor: Fast, Secure, and Programmable
Subprocess Virtualization}

\author{
{\rm Bumjin Im}\\
Rice University
\and
{\rm Fangfei Yang}\\
Rice University
\and
{\rm Chia-Che Tsai}\\
Texas A\&M University
\and
{\rm Michael LeMay}\\
Intel Labs
\and
{\rm Anjo Vahldiek-Oberwagner}\\
Intel Labs
\and
{\rm Nathan Dautenhahn}\\
Rice University
} % end author

\hypersetup{draft}
\date{\vspace{-3em}}
\begin{document}
\setlength{\droptitle}{-5em}
%%----------------------------------------------------------

\maketitle

\input{abstract}
\input{intro}
\input{motivation}

%\input{bg}
\input{arch}

\input{design}

\input{usecase}
\input{eval}

%\input{disc}
\input{rw}
\input{conc}
%\input{acks}
%\input{avail}

%-----------------------------------
\printbibliography

%-----------------------------------
\appendix
\input{mpksep-bg}

\input{sec_eval_apdx.tex}
\input{ceteval}

\input{app-32bit-mode}

\input{archive-flow}

\input{app-sig}

%%%%%%%%%%%%%%%%%%%%%%%%%%%%%%%%%%%%%%%%%%%%%%%%%%%%%%%%%%%%%%%%%%%%%%%%%%%%%%%%
\end{document}

%% file: header.tex
\usepackage[
  backend=biber,
  style=numeric-comp,
  maxbibnames=99,
  url=true, doi=false,isbn=false,
  citestyle=numeric-comp
]{biblatex}
\AtEveryBibitem{
    \clearfield{urlyear}
    \clearfield{urlmonth}
}
\usepackage{listings}
\lstdefinestyle{customc}{
  belowcaptionskip=1\baselineskip,
  breaklines=true,
  language=C,
  showstringspaces=false,
  basicstyle=\footnotesize\ttfamily,
  keywordstyle=\bfseries\color{green!40!black},
  commentstyle=\itshape\color{purple!40!black},
  identifierstyle=\color{blue},
  stringstyle=\color{orange},
}

\microtypecontext{spacing=nonfrench}

\usepackage{times}
\usepackage{longtable}
\usepackage{multirow}
\usepackage{wrapfig}
\usepackage{amsmath, amsthm, amssymb}
\usepackage{amsmath}
\usepackage{wasysym} % moon symbols
\usepackage{amsfonts}
\usepackage{fancyhdr}
\usepackage{graphicx}
\usepackage{textcomp}
\usepackage{floatrow}
\usepackage{tabularx}
\definecolor{riceblue}{RGB}{0,32,91}

\usepackage{xspace}
\usepackage{rotating}
\usepackage{tabularx}
\usepackage{booktabs}

\newcommand{\etal}{\emph{et al.}\xspace}
\newcommand{\eg}{\emph{e.g.,}\xspace}
\newcommand{\ie}{\emph{i.e.,}\xspace}
\newcommand{\etc}{\emph{etc.}\xspace}

\newenvironment{possiblecut}{}{\ignorespacesafterend}

\ifdefined\docuts
  \renewenvironment{possiblecut}{\expandafter\comment}{\expandafter\endcomment\ignorespacesafterend}
\fi

\ifdefined\showcuts
  
\fi

\newcommand{\Code}[1]{\mbox{\texttt{#1}\xspace}}

\usepackage[compact]{titlesec}
\titlespacing{\paragraph}{0pt}{1pt}{8pt}

\newcommand{\intel}{Intel\textsuperscript{\textregistered}\xspace}

\newif\ifcomments
\commentstrue
\ifcomments
  \newcommand{\review}[1]{{\bf\textcolor{red}{[#1]}}}
  \newcommand{\notify}[1]{{\bf\textcolor{blue}{
    [COMMENT FOR EDITOR -- #1]
  }}}
  \newcommand{\notes}[1]{{\footnotesize\textcolor{green}{[NOTE: #1]}}}
  \newcommand{\writeme}[1]{{\textcolor{gray}{[#1]}}}
  \newcommand{\todo}[1]{{{\it \color{red}\{#1\}}}}

  \usepackage{xcolor}
  \definecolor{darkgreen}{rgb}{0,0.6,0}
  \newcommand{\ndd}[1]{\fcolorbox{cyan}{white}{}\footnote{\textcolor{cyan}{[#1--ndd]}}}
  
  \newcommand{\structure}[1]{\noindent{\em\textcolor{darkgreen}{[#1]}}}
  \newcommand{\point}[1]{{\bf\textcolor{violet}{<#1>}}}
  \newcommand{\paroutline}[2]{\textbf{\structure{#1}\point{#2}}}
  \def\outline#1{}
  \newcommand{\secpurpose}[1]{{{\em\textcolor{gray}{Section Purpose: #1}}}}
\else
  \def\ndd[1]{}
  \def\review[1]{\errmessage{NO review macros when comments turned off!}}
    \def\notify[1]{}
    \def\notes[1]{}
    \def\writeme[1]{}
    \def\todo[1]{}
  
  \def\structure{}
  \def\point{}
  \newcommand{\paroutline}[2]{}
  \def\secpurpose#1{}
\fi

%% file: abstract.tex
%%----------------------------------------------------------
\begin{abstract}
%%----------------------------------------------------------

  % --------------------
  Commodity applications contain more and more combinations of interacting
  components (user, application, library, and system) and exhibit increasingly
  diverse tradeoffs between isolation, performance, and programmability.
  %
  %Unfortunately, due to complexity and performance they use monolithic process
  %based isolation, resulting in reliability and security problems, while bolt-on
  %security frameworks are rigid and exploitable.
  % --------------------
  We argue that the challenge of future runtime isolation is best met by
  embracing the multi-principle nature of applications, rethinking process
  architecture for fast and extensible intra-process isolation.
  We present, the \visor, a new process model and security architecture that
  nests an extensible monitor into the standard process for building efficient
  least-authority abstractions.
  The \visor introduces a new virtual machine abstraction for representing
  subprocess authority, which is enforced by an efficient self-isolating monitor
  that maps the abstraction to system level objects (processes, threads, files,
  and signals).
  % --------------------
  We show how the \arch can be used to develop specialized separation
  abstractions using an exokernel-like organization to provide virtual privilege
  rings, which we use to reorganize and secure NGINX.
  Our prototype, includes a new syscall monitor, the \emph{\nexpoline}, and
  explores the tradeoffs of implementing it with diverse mechanisms, including
  \intel Control Enhancement Technology.
  Overall, we believe sub-process isolation is a must and that the \arch exposes
  an essential set of abstractions for realizing this in a simple and feasible
  way.

\end{abstract}
%%----------------------------------------------------------

%% file: intro.tex
%%----------------------------------------------------------
\section{Introduction}
%%----------------------------------------------------------

  %% Point: Process based separation is inflexible and does not match
  %% multiprincple
  % The core issue is that the process is the de facto representation for access
  % control but processes house many principles in practice.
  % -------------------- process is
  The \emph{process} abstraction~\cite{Ritchie74Unix} defines an interface between an application
  and system resources, which is heavyweight and slow for cross-component interactions.
  %% process doesn't adequately support today's apps
  %Unfortunately, this interface is heavyweight and significantly slows
  %interactions between components, leading to
  However, without process isolation, diverse multi-principle
  environments with large fault domains will suffer reliability and security
  problems~\cite{Durumeric:Matter:2014, Hao02Setuid}.
  %% workarounds are inflexible
  As a result, many attempts have been made to control access to system
  resources (files, address spaces, signals) through system object virtualization~\cite{Engler95Exokernel,Porter11Drawbridge,Madhavapeddy13Unikernels,Tsai:Cooperation:2014}
  or kernel-level mechanisms~\cite{Soltesz07Container,Price04SolarisZone,LXC,Zeldovich11HiStar,Nikolaev13VirtuOS}.
  %and integrating access control frameworks into the kernel.
  %
  But, all of these languages enforce rigid interfaces that, while quite
  expressive, are not extensible and by nature introduce threats to
  unrelated parts of the kernel.
  Beyond that, they also fail to monitor and enforce resource control at the
  subprocess level.
  %
  %Alternatively, many have attempted to split an application across multiple
  %processes, however, this suffers from prohibitive slowdowns.
  %
  % missing part of arg: apps require custom abstractions: we need to motivate
  % extensibility and customizability.

  % --------------------
  %We believe the best way forward is to think about security frameworks fitted
  %in between the OS and application that can isolate inside the process.

  \iffalse
  % These are origianl sentence starters for the first paragraph, in an effort
  % to not boggle my mind trying to fit them in, just comment out for now and
  % get a complete draft up.
  % --------------------
  The process abstraction doesn’t work for security principles despite 50 years
  trying.
  %

  % --------------------
  The process is archaic, and while still useful, is unsuitable for today's
  commodity applications that combine diverse runtime elements, languages, and
  libraries.

  % --------------------
  Too long has the security community relied on the \emph{process} abstraction
  for enforcing access control.
  %
  The antiquated nature forces bolt on security models in the kernel and large
  scale distributed frameworks on top of it, so that isolation can be had.

  \fi

  %% Point: IPI attempts but lacks several elements
  % --------------------
  Until recently, software fault isolation~\cite{Wahbe:Efficient:1993, Castro09SFI, Sehr10SFI, Mao11SFI} was the only practical method for
  isolating elements within a single address space, which uses inline reference
  monitors.
  Fortunately, both hardware and software vendors have observed this trend and
  invested into subprocess isolation.
  % Chia-Che: To me, it's incorrect to say these hardware/software isolation lacks OS support, since many have been embraced by the Linux community.
  %, despite the lack of OS support.
  %
  These systems include memory and CPU virtualization that allows for multiple
  protection domains within a process with fast domain switching
  (memory protection keys~\cite{Park:libmpk:2019}, nested paging fast context-switch
  support~\cite{intel-vt,amd-v}, etc.).
  The core technique is to nest a monitor into the process that efficiently
  enforces subprocess access control to memory and CPU
  state~\cite{Vahldiek-Oberwagner:ERIM:2019,Hedayati:Hodor:2019,Hsu:Enforcing:2016,Chen:Shreds:2016,CheriABI}.%\ndd{secage doesn't have this
  %limitation? maybe its not extensible or lacks system resource control?}
  %
  Unfortunately, these approaches only virtualize minimal parts of the CPU and
  neglect tying their domain abstraction to system resources.
  % Chia-Che: already listed in paragraph 1
  %concurrency,
  %address spaces, files, and signals.
  %
  %Unfortunately, prior work only monitors memory access, which is only one way
  %to access information in a process, \ie data is accessible through system
  %objects and CPU registers, which can be used by attackers to bypass privilege
  %separation~\cite{Connor:PKU:2020}.
  %
  %These flows result from 1) the lack of a \subdomain abstraction in the
  %kernel, leading to leaky system objects, and 2) missing abstractions in the
  %\nemo mechanisms, leading to low-level exploits.
  %
  %First, system objects allow bypass through signals, \syscalls, and forked
  %address-spaces.
  %
  As a result attackers can easily break out of the sandbox.
  For example, suppose \Code{s} points at an isolated session key, then the
  following pseudo code would bypass any memory protection:
  % --------------------
  \begin{lstlisting}[language=C,style=customc]
  open("/proc/self/mem","r").seek(s).read(0x10) \end{lstlisting}
  %
  %Signals expose protection information in virtual CPU state (\eg
  %\Code{pt\_regs}) and can be used to coerce control and data flow.
  %
  %Second, prior work neglects full implementation and analysis of
  %multi-threaded concurrency and multi-domain separation, which means several
  %access paths are left undefended: without implementing concurrency, \monitor
  %state can be corrupted or security checks bypassed and in multi-domain the
  %context switch exposes attacker controllable registers.
  %

  Alternatively, other intra-process isolation approaches, like Native Client~\cite{Yee:Native:2009},
  avoid this problem by completely eliminating access. However, this severely
  limits functionality of the code, and neglects a large application space.
  It is clear that system objects require more than all or nothing access, which
  says nothing about signals and IPC.
  %
  %Table~\ref{tbl:barf} and Figure~\ref{fig:problem} summarize key differences
  %and limitations of prior work.  % removed with table 1
  %
  Beyond these security gaps, these approaches %are point solutions representing
  made ad hoc guesses at the best securing abstractions,
  but do not provide detailed specialized protection actually useful for applications.

  % --------------------
  \begin{figure}[t]
      \begin{center} \includegraphics[width=.9\textwidth]{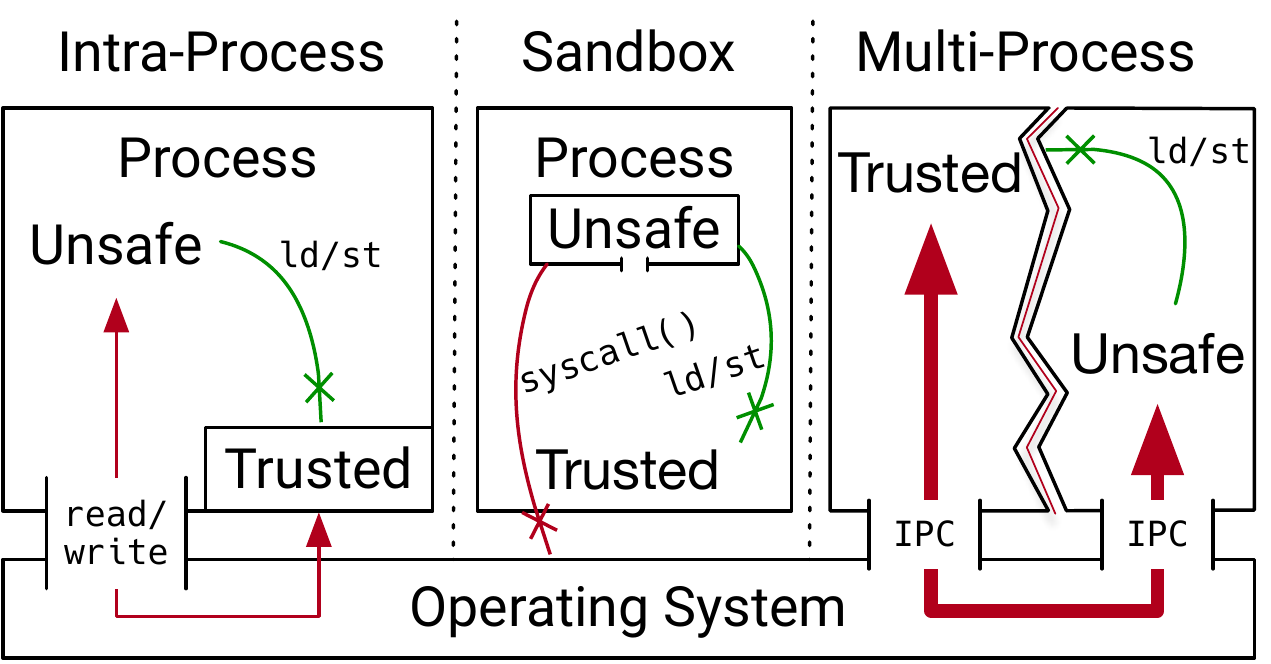}
      \end{center}
      \caption{\label{fig:problem}
        Problem: intra-process is bypassable because domain is opaque to OS,
        sandbox limits functionality, and inter-process is slow and costly to
        apply. Red indicates limitations.
      } \vspace{-3mm}
  \end{figure}

  \iffalse
  %
  After decades of research, IPI has demonstrated several valuable systems, but
  remains as a set of point solutions that while improving on efficiency of
  enforcing IPI with process level isolation or limiting functionality for fast
  single-address space IPI, remain fixed and lack true isolation for the runtime
  (as exemplified by recent exploits).
  %
  After investigation, we also found state of the art to neglect key operating
  systems challenges such as mapping low-level memory isolation to the system
  abstractions (process and thread), concurrency, signals, and exploitability in
  existing mechanisms.
  %
  It is clear that the trend towards IPI single address space isolation is not
  an enthusiast only mission, as major hardware vendors have explored many
  different types of mechansism released into production.
  %
  \fi

  %Our goal is to preserve the speed of nested privilege separation while making
  %it secure, fully supportive of system interfaces, and imminently deployable.
  % --------------------
  We believe that the core problem is the disconnect between the need for
  subprocess level resource control and the inability of existing solutions to
  extensibly and fully abstract it.
  The core research question of this paper is: \emph{Can we easily add an efficient and
  extensible security architecture that does not decrease
  the security of the OS while fully virtualizing access to all
  system resources?}
  % --------------------
  %This paper, leverages new production IPI mechanisms called memory protection
  %keys to build a new extensible OS architecture that resides solely in users
  %space.
  As such, we explore a new process model, the \emph{\arch}\footnote{endo-:
  greek for ``within`` and stands in contrast to exo-}, where the monitor is
  nested within the process to export a lightweight virtual machine for
  representing and enforcing subprocess resource isolation
  %
  %The key idea is to nest the monitor into the process, develop a novel
  %lightweight VM abstraction by which to specify policies in, and then to
  %develop extensible interfaces for support
  with novel custom security languages fit
  for the application at hand.
  There are two key innovations: 1) We map the new virtual machine to the
  traditional process level abstractions to ensure no leaks that plagued the prior work, and 2) make it efficient and extensible to
  support custom protection abstractions. % Chia-Che: redudant! % using the basic virtual machine.
  We demonstrate the value of \arch by developing a simple yet powerful virtual
  ringed process architecture to address several common exploit
  patterns(\S\ref{sec:gaps}).

  \iffalse
  % --------------------
  %Beyond the process abstractions, we must deal with several challenges of
  %deploying a nested security monitor.
  %
  Our prototype, \system, addresses several unsolved challenges in the
  implementations of prior systems and explores tradeoffs for system call
  monitoring mechanisms.
  %
  Namely, prior work neglected to implement concurrency, true multidomain within
  a process and across threads, and properly virtualize signals.
  %
  We also introduce a unique method for monitoring system interactions with the
  \nexpoline and novel exploration of new hardware for optimizing control flow
  integrity (\intel CET).
  %
  \fi

  % --------------------
  Our prototype \visor, named \emph{\system}, uses memory protection keys for efficient
  memory isolation, with a new monitor to intercept all \syscalls and virtualize
  signals, and is applied to implement and evaluate concurrency and multi-domain
  mechanisms.%\ndd{this paragraph needs a reduce pass, too low-level now.}
  % Chia-Che: This is my attempt, take a look
  %
  %Key insights include introducing new \syscall monitoring and signal
  %virtualization techniques.
  %
  The monitor enforces using a new secure \syscall callgate, \emph{\nexpoline}, which we implement using several diverse low-level mechanisms
  %and evaluate performance vs security tradeoffs.
  %
  %The key challenge is
  to ensure that only the monitor is permitted to invoke a
  \syscall.
  To virtualize signals, \system inserts a layer of indirection and maps signals
  to the appropriate \subdomain, and has to overcome
  %which is challenging because of the added
  the complexity from concurrency and the subtle, poor interfaces
  presented by the kernel.
  \system also identifies and prevents multi-process attacks that can bypass
  policies via copy-on-write address spaces.
  % Chia-Che: The following description is redundant.
  % --------------------
  %We present the first implementation of fine-grained concurrency management for
  %mutli-threaded \monitor runtime, which is complex because of subtle TOCTOU
  %vulnerabilities and performance implications.
  %
  Last, while prior work indicated multi-domain separation, we identify
  unexplored low-level threats due to the domain switching mechanism, which we
  address with a secure context-switch gate.

  We evaluate and compare \system to a ptrace-based approach and
  ERIM~\cite{Vahldiek-Oberwagner:ERIM:2019} which provides only memory
  protection but no system isolation.
  We also implement and apply several exploits to test the security of our
  solution.
  Overall, overheads range from 2-40\% while closing many security gaps.
  %
  % --------------------
  In summary our contributions include:
  %
  %\begin{itemize}[leftmargin=13pt]
  \begin{itemize}
    \item New architecture for \subdomain virtualization: \visor architecture
      that virtualizes system objects solely at user level while remaining fast
      and securing against new low-level attacks (\S\ref{sec:arch} and
      \S\ref{sec:design}).
    \item A prototype, \system, that includes concurrency and scalability
      (\S\ref{sec:design} and \S\ref{sec:eval}); the \nexpoline definition and
      exploration of several lightweight \syscall monitoring mechanisms
      (\S\ref{sec:syscallvirt}), including a novel use of \intel Control
      Enforcement Technology (CET); and a systematic represenatation of system
      object virtualization (\S\ref{sec:sysflow}).
    \item Programmable security abstractions (\S\ref{sec:programmable}) with
      example \Boxing facilities (\S\ref{sec:boxing}) and novel decomposition of
      \Code{sudo} and NGINX (\S\ref{sec:usecase}), demonstrates only 5.5\%
      overhead for securing private keys and preventing parser-bug privilege
      escalation (\S~\ref{sec:boxeval}).
  \end{itemize}

%% file: motivation.tex
%%----------------------------------------------------------
%\section{The Problem of Nested Process Virtualization}
%\section{Background and Related Work}
%%----------------------------------------------------------
\section{Motivations}\label{sec:motivation}
%%----------------------------------------------------------

  %\subsection{Application Monoculture}
  %\subsection{Inflexible Security Architectures}

  % --------------------
  Modularity is traditionally provided using virtualization, where the hardware,
  OS, and programming languages combine to provide the illusion that a thread of
  execution has the full machine.
  The basic abstraction for virtualization is a process, where one or multiple threads of
  execution share an address space and virtualized system objects.
  The problem is, a process is heavyweight and execution in the same address space tends to fail to apply least-authority
  design.
  As a result,
  malicious or compromised components will allow
  an attacker access to the full authority of the process.

  % --------------------
  In this section we argue that the challenges facing future application
  resource control are incompatible with the existing process-based isolation. % and lay out a
  %set of goals and challenges we anticipate are necessary for handling the
  %future of application resource management.
  %
  While these application trends are not new~\cite{efstathopoulos2005labels, Adrian10JoeE, Bittau:Wedge:2008a,TaMin06Splitting}, innovations
  in hardware and software fault monitors along with nested resource
  virtualization open up a new opportunity for fast, safe, and extensible
  process virtualization.
  %
  %At our core we argue that the future of application resource management
  %requires the following: 1) full virtualizability, 2) extensible and
  %programmable abstractions, 3) lightweight abstractions with bind time
  %authorization and lightweight checks, and 4) complete mechanisms.
  % --------------------
  We argue that future application resource management must address the
  following problems: 1) lack of full virtualizability leads to incompatibility and
  insecurity, 2) lack of extensibility leads to hard-to-apply abstractions, 3)
  incomplete implementations lead to insecurity and inconclusive
  results, and 4) lack of hardware portability complicates integration. % and 4) lack of a hardware portability integrates mechanisms and
  %policy limiting systematic consideration.

  % --------------------
  \subsection{Full Virtualizability}
  % --------------------

    % --------------------
    Just like the original challenge of VMMs set out by Popek and
    Golberg~\cite{Popek74}, current subprocess isolation suffers from incomplete
    virtualization.
    To complete subprocess isolation,
    we need to create a virtual machine that
    is a subset of the standard process while also multiplexing OS objects.
    %
    %The goal is to multiplex the single process resources, however, many
    %elements are not virtualized.
    %
    Incomplete virtualization leads to direct access that breaks the isolation.
    Problematically, most systems that provide this type of isolation focus
    primarily on isolating memory but neglect control flow and data access that
    are made available through the system resources,
    including file system and address space objects.
    Additionally, interrupted program state and signal delivery are neglected
    and challenging to support.

    % --------------------
    On the other hand, some systems apply proper sandboxing, such as Native
    Client~\cite{Yee:Native:2009}, but only provide restrictive virtual machine abstractions that fail to implement the
    desirable applications.
    For example, the sandboxed code cannot make system calls,
    %
    %Current subprocess isolation approaches are all or nothing sandboxes, which
    %is powerful but loses functiionality.
    %
    %Many applications demand partial access: for example, \todo.
    %
    effectively providing only all or nothing policies. %virtualization, however, many
    %applications needs partial.
    %
    Sometimes, parts of an application may need access to some but not all
    files, such as access to secret and sensitive data, while other components need access to other, less-sensitive data,
    such as configuration files.
    Existing interfaces for OS object virtualization are also insufficient for supporting more flexible policies at subprocess level.

  % --------------------
  \subsection{Programmable Security Abstractions}
  % --------------------

    % --------------------
    Much like the Exokernel argument~\cite{Engler95Exokernel}, today's process-based isolation is
    inflexible.
    However, unlike Exokernel, the key challenge is not about exposing state for
    managing performance, but rather making the policy language more closely
    matching the needs of applications.
    This influences 1) Ease of use:
    A primary reason why fine-grained security is not applied is the
    complexity and diverse nature of application demands.
    % --------------------
    We argue that an abstraction that works for one application won't
    necessarily be the easiest to apply for another.
    2) Performance. %Additionally, the abstraction impacts performance.
    % --------------------
    We believe that an extensible protection architecture will ameliorate these
    issues by putting control into application specific abstractions.

  % --------------------
  \subsection{Mechanism Portability}
  % --------------------

    % --------------------
    The key problem is
    what are the essential elements independent of the mechanisms.
    It is clear that intra-process isolation mechanisms are only going to see
    increased exploration, which fractures the landscape of approaches for
    applying them.
    Each new system provides some properties, but how do we compare them?
    We believe it is necessary to establish a model that prescribes a set of
    clear abstractions and security properties so that diverse systems can be
    reasonably and systematically applied and compared.

  % --------------------
  %\subsection{Incomplete Implementations}
  % --------------------
    \input{bg}

%% file: bg.tex
%%----------------------------------------------------------
%\section{The Problem of Nested Privilege Separation}
%\section{Background and Related Work}
%\section{Gaps and Challenges}
\subsection{Mechanisms Gaps and Challenges}
\label{sec:gaps}
    Several facets must be preserved to have % removed with table 1 
    meaningful privilege separation and compares related efforts.
    The key gaps and challenges are described below.

    % Arg: IPI must be extended to consider these threats/problems 
    % --------------------
    %We define problems with existing work as malicious information flows that
    %are made available to an attacker through objects within the runtime
    %enviornment.
    %
    %We first detail the key missing security ingredients, then describe
    %challenges in bridging the gap to a complete solution.
    %
    %Although existing work isolated per thread regions, they fail adequeately
    %addres severl issues from... 
    %
    %The key missing resources and abstractions include: system objects as made
    %available through the system call interface, signal abstraction which
    %influence both correctness of control flow and data integrity of critical
    %register state, CPU registers that are accessible via ..., ..., ...,

    %A solution requires isolation of syscalls, signals, memory, cpu registers,
    %and control flow.
    %

    %%--------------------------------------------------------
    \paragraph{Subdomain Identifiability}
    %%--------------------------------------------------------
      One solution would be to extend the kernel with subprocess abstractions. However,
      a userspace monitor is still necessary to track
      the current protection domain or else you have to
      transition to kernel on each switch which is
      prohibitively costly.
    
    %%--------------------------------------------------------
    \paragraph{Programmability and Optimizations}
    %%--------------------------------------------------------
      % --------------------
      Having a general interfaces would be ideal but as implored by prior
      work (Exokernel, etc), applications tend to be severely constrained.
      What's worse is that existing process abstraction needs to have a separate interface to accommodate different interaction pattern and to be efficient. 
      %This problem is made worse when thinking about
      %separating existing processes where interfaces are
      %irregular and interaction patterns might require
      %different separation abstractions to be efficient.
      %
      Thus custimizeability of the abstractions is critical
      and most prior work don't handle it properly.

    %%--------------------------------------------------------
    \paragraph{Leaky System Objects}
    %%--------------------------------------------------------

    \iffalse
      % --------------------
      %Unfortunately, the isolation properties of a \nemo---control and data
      %integrity---can be bypassed through system calls or signal manipulation.
      %
      %Connor \etal analyzed MPK based systems found several attacks that disable
      %or bypass protections~\cite{Connor:PKU:2020}.
      %
      The OS does not respect MPK domains, and as such has several interfaces
      that expose isolated data (detailed assessment and model of policies
      presented in \S~\ref{sec:sysflow}).
      %
      %Any intra-privilege protection system must prevent these threat vectors,
      %otherwise all protection is trivially bypassed.
      %
      Sirius~\cite{Tarkhani:EnclaveAware:2020} is the only solution that considers
      this threat but proposes to integrate into the kernel itself, which
      results in a complex implementation with enforcement checks scattered
      throughout the kernel.  
      %requiring the kernel adhere to a novel intra-process abstraction.
      %
      Instead, we use the system call interface to check and enforce intra-app
      privilege separation within a single layer.
      %
      A key question is what is the set of minimal and comprehensive policies to
      eliminate the threat?
      %
    \fi
  
      % --------------------
      %\paroutline{FundamentalProb}{OS Misses MPK} 
      % --------------------
      Since OSs are unaware of subprocess domains, an untrusted
      portion of a application can request access and the OS will
      gladly service it.
      %---its security principles are at the application and user granularity,
      %
      %For example, suppose \Code{s} points at an isolated session key, then the
      %following pseudo code would bypass any memory protection:
      % --------------------
      %\begin{lstlisting}[language=C,style=customc]
      %open("/proc/self/mem","r").seek(s).read(0x10) \end{lstlisting}
      %
      Although we shows several bypass attacks, the primary challenge is to systematically
      assesses all interfaces and to integrate them into a unified policy management
      interface.
      %
      %Beyond that, Paper exposed a fundamental problem with all \nemo, the OS
      %can do wahtebver it wants. 
      %
      It is easier to reason about the policy for a relatively strict interface, but things like ioctls make it impossible to have comprehensive defenses.

    %%--------------------------------------------------------
    \paragraph{System Flow Policies}
    %%-----------------------------
      % --------------------
      A basic property is that the information in a \segment should never flows in or out of
      system objects unless explicitly granted. 
      % --------------------
      However, deriving the system flows itself is hard due to system complexity.  
      %
      %Why? 
      %
      %complexity.  
      Although prior work such as Erim and Hodor shows that
      one can reason about the flows through a specific system object, the approach is hard to be broadened to a systematic solution.
      %While prior work showed bla, they were not systematic and
      %focused too much on narrow erim/hodor.
    
    %%--------------------------------------------------------
    \paragraph{\syscall Monitor}
    %%-----------------------------
      % --------------------
      The need to monitor \syscalls is clear, but how to do it is not.
      %
      %The kernel could be modified to properly abstract \domains, however, 1)
      %it is unlikely to gain traction within the Linux community, and 2) more
      %significantly, the complexity of these policies inside the kernel is
      %extensive.
      %
      %Analyzing and specifying policies at the system call abstraction level
      %allows for a \syscall only monitor to enforce the policies.
      %
      %The challenge is in controlling access to \syscall instructions so that
      %all instances go through the \monitor.
      % 
      A deny-all policy---as used by intra-app
      sandboxing~\cite{Yee:Native:2009, WebAssemblyCommunity:Security:2020,
      Narayan:Retrofitting:2020}---sandboxing would indiscriminately deny all access and neglect a
      large application space.
      For example, deny-all sandboxing cannot prevent
      Heartbleed~\cite{Durumeric:Matter:2014}. 
      In general, \emph{applications should be able to benefit from privilege
      separation while not losing functionality.}
      Alternatively, we could modify the OS so that it recognizes and enforces
      \subdomains~\cite{Litton:Lightweight:2016, Hsu:Enforcing:2016,
      Tarkhani:EnclaveAware:2020}.
      Unfortunately, this introduces significant complexity as indicated by
      Sirius~\cite{Tarkhani:EnclaveAware:2020}.
      %which modified 15k LoCs to enforce \segment isolation
      %policies~\cite{Tarkhani:EnclaveAware:2020}. 
      %
     % Moreover, ad hoc integration creates the potential many more gaps.
      %
      %Linux has consistently opposed any notion of intra-process separation
      %abstractions due to its high complexity and requiring checks and
      %enforcement throughout the kernel.

      % --------------------
      %\paroutline{In-Kernel}{Problems with inkernel} 
      %
      Instead of the in-kernel approach, we propose
      enforcing nested flow policies at the \syscalls---allowing some to bypass
      without change, others to be denied, and the rest to be securely emulated.
      This is not supported by well-known systems in Linux:
      MBOX uses ptrace for similar protections~\cite{Kim:Practical:2013},
      but only virtualizes the filesystem interface and is inefficient.
      Seccomp~\cite{Corbet:Seccomp:2009} with
      eBPF~\cite{Fleming:thorough:2017} and LSM~\cite{Wright:Linux:2003} enforce \syscall policies, but lack the ability to
      modify \syscall semantics, which will require modifying the LSM hooks extensively.
      %virtualize \syscall: they can only deny or accept.
      %
      %To properly virtualize \syscalls, the LSM hooks would have to be vastly
      %modified to allow modification of arguments and splitting \syscalls.

    %%--------------------------------------------------------
    \paragraph{Multi-Process}
    %%--------------------------------------------------------
      % --------------------
      An attacker can fork an exploited process, and access the original address
      space directly through load and stores instructions and access indirectly
      through read system calls. % to memory in the original process.
      The \monitor must be inside the new process to ensure the protections, or
      the memory must be scrubbed.
      Prior approaches~\cite{Tarkhani:EnclaveAware:2020,
      Vahldiek-Oberwagner:ERIM:2019, Hedayati:Hodor:2019} do not consider this
      threat and would have to disallow \fork system calls.
      
    %%--------------------------------------------------------
    \paragraph{Signals}
    %%-----------------------------
      % --------------------
      Signals create several exploitable gaps and challenges.
      %
      %can coercing control flow state by specially timed interrupts, thus
      %giving attackers control.  , but a more subtle issue is that Linux hard
      %codes many elements of their processing, which creates several problems.
      %
      First, Linux exposes virtual CPU state to the signal handler including
      \pkru, which can be exploited by an attacker.
      Second, the kernel does not change the domain and will trap if not
      properly setup. 
      Third, the kernel always delivers the signal to a default domain, exposing
      the monitor control-flow attacks.
      Fourth, properly virtualizing signals requires complex synchronization and
      modifying the semantics to be both correct, safe, and efficient.
      Overall, properly handling signals introduces significant complexity into
      the \monitor.

    %%--------------------------------------------------------
    \paragraph{Multi-Threaded}
    %%--------------------------------------------------------
      % --------------------
      While existing work claims to have a design supporting multi-threading,
      none of them have implemented concurrency control in the runtime monitor,
      introducing TOCTOU attacks and memory leaks, as well as neglects to
      measure scalability.
      %
      %The complexity in order to both properly protect \monitor data structures
      %as well as make the runtime efficient requires significant development.
    
    %%--------------------------------------------------------
    \paragraph{Multi-Domain}
    %%--------------------------------------------------------
      % --------------------
      Prior work isolated one domain per thread but not multiple domains per
      thread.
      The challenge is that switching from the \ud to the monitor exposes less
      data than executing an cross domain call because the stack requires
      tracking to ensure return integrity.

%% file: arch.tex
%%----------------------------------------------------------
\section{\model}
\label{sec:arch}
%%----------------------------------------------------------

  % --------------------
  %\begin{figure*}[]
  \begin{figure}[]
      \begin{center}
        \includegraphics[width=\textwidth]{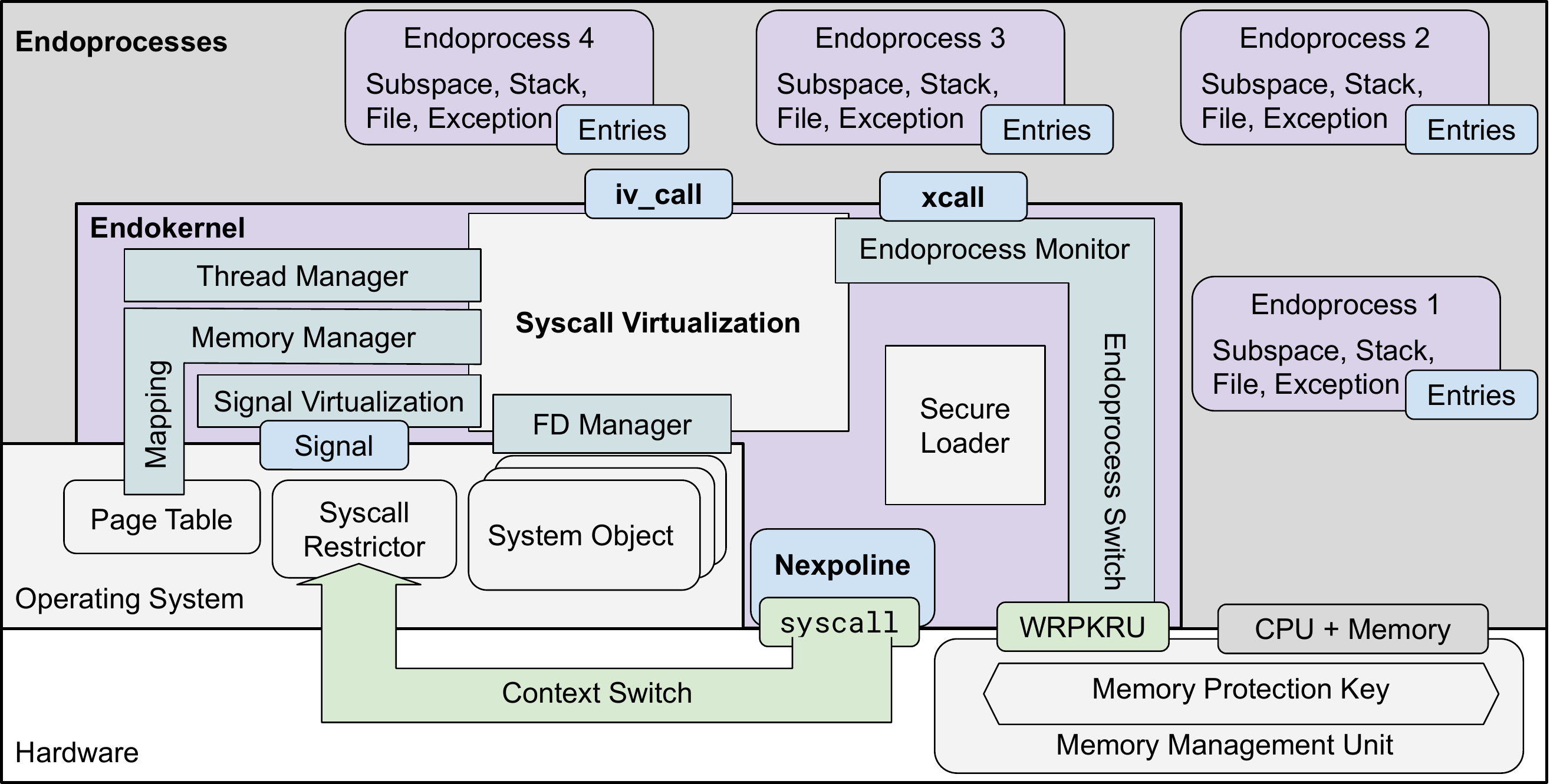}
      \end{center}
      \caption{\label{fig:sysarch}
        \system Architecture.
      }
  %
  %\end{figure*}
  \end{figure}

  % --------------------
  The \model is a general purpose model for nesting a monitor, the
  \emph{\monitor}, into the process address space, that self-isolates to
  create two privilege levels within the process, and presents a lightweight
  virtual machine, the \subdomain, to the application.
  The \visor has been designed to insert directly below application logic and
  directly on top of the OS and HW provided abstractions.
  The core methodology is to systematically identify 1) what needs to be
  protected, 2) how that information can be interacted with (through the CPU,
  memory, or OS interfaces), and 3) specify a set of abstractions that must be
  in place to secure \subdomain isolation.
  The basic goal is to identify an architecture level description that is
  portable and independent of the exact layers above and below to properly
  encapsulate the \subdomain internals.
  The architecture has two main elements: 1) the authority model and 2) the
  nested \monitor architecture that ensures isolation. 
  Figure~\ref{fig:sysarch}, depicts these three elements together in the
  architecture.

  \iffalse
  %
  Our goal is to provide separation of policy and mechanism so that 1) the
  architecture is portable to many mechanisms and 2) diverse implementations can
  be explored.
  %
  We first present the authority model, then use that to describe the basic
  \visor organization and associated policies, and then describe a custom
  abstraction, \emph{nested boxing}, for applying least-authority policies.
  \fi

  \iffalse
  % --------------------
  One of our core insights is to recognize resources that standard nested
  separation systems neglect and make them first class abstractions in the
  protection system. This allows us to reason about and specify policies for
  access. Our goal is first to deny any circumvention of the nested separation
  monitor that resides within the address space as the untrusted part. As
  indicated by our bg this means we need to deal with syscalls, signals, and ...
  --

  \fi

  % --------------------
  \input{principles}

  % --------------------
  \input{privmodel}

  % --------------------
  \input{org}

  % --------------------
  %\input{properties}

%% file: principles.tex
  %%--------------------------------------------------------
  \subsection{Design Principles}
  \label{sec:principles}
  %%--------------------------------------------------------

    % --------------------
    We share the trusted monitor principles as outlined by
    Needham---tamper-proof, non-bypassable, and small enough to
    verify~\cite{needham1972protection}---and add the following:
    % --------------------
    \paragraph{Nested Separation Kernel}
    Address spaces and kernel interactions are slow, eliminate all OS
    interactions~\cite{Rushby:Design:1981, Dautenhahn:Nested:2015}---\ie pure
    userspace, while being smaller than a microkernel and only tolerating
    elements inside if they support primitive separation mechanisms with a
    minimal interface. 

    \iffalse
    % --------------------
    \paragraph{Microkernel Influenced}
    %
    \emph{A concept is tolerated inside the microkernel only if moving it
      outside the kernel, \ie permitting competing implementations, would
      prevent implementation of the system’s required
      functionality~\cite{Liedtke:Microkernel:1995}.}

    % --------------------
    \paragraph{Minimal Interface}
    %
    The likelihood for mistakes increases as complexity increases, therefore,
    the monitor must expose the most limited interface to the untrusted
    world---this manifests in how much functionality is inserted into the
    monitor: \ie no glibc code.
    %
    \fi

    \iffalse
    % --------------------
    \paragraph{No Global CFI}
    %
    Enforcing coarse grained CFI~\cite{Abadi:Control:2007} simplifies isolation
    but can significantly degrade performance.
    %
    \fi

    % --------------------
    \paragraph{Self-Contained and Secure Userspace}
    Avoid implementing system object isolation in the kernel: adding yet another
    security framework hacked on top of thousands of kernel objects.  
    %
    %Instead, rely on protection policies at the \syscall interface.  
    %
    Nesting requires part of the mechanism to be in-process, however, certain
    resources could be virtualized by the OS. 
    While that seems like it might be the best choice, if parts of the process
    were virtualized by the monitor and others by the OS then: 1) complexity
    arises in bridging the semantic gap of the abstractions, 2) bugs can arise
    from complex concurrency, access, and exception control, and 3) ties the
    \subdomain abstraction to a specific kernel implementation instead of the
    semantics of its interface.
    %
    %Furthermore, we anticipate this architecture would be deployable in IoT or
    %non-standard environments. 
    %
    %We argue for encapsulating \subdomain isolation within a complete and single
    %monitor component using the principle of \emph{monitor independence}. 

    % --------------------
    \paragraph{General and Extensible}
    The design should permit many implementations, \ie using various hardware
    (MPK) or software isolation (SFI) techniques that might present valuable
    tradeoff points in the security-performance space.
    The architecture should enable safe extensibility of the security
    abstractions to enable custom, least-authority protection services.

%% file: privmodel.tex
%%--------------------------------------------------------
\subsection{Authority Model}
\label{sec:privmodel}
%%--------------------------------------------------------

  % --------------------
  The \visor represents and enforces authority based on a protection \domain,
  called an \emph{\subdomain}.
  As outlined by Lampson~\cite{Lampson:Protection:1974} and instantiated by
  Mondrix~\cite{Witchel:Mondrix:2005}, an \subdomain must provide the basic
  properties of data abstraction: protected entry/return and memory isolation,
  while also protecting access through OS objects.
  Most existing work multiplexes regions of the virtual address space and
  protects control flow, however, these works neglect to map memory space
  isolation to the system objects. 
  Thus, in addition to CPU and memory virtualization, the \visor also
  virtualizes: CPU registers, the file system, mappings, and exceptions (as
  implemented through signals).
  %
  %The \emph{authority context} is a lightweight \subdomain virtual machine.
  % --------------------
  %The basic protection context, \emph{\subdomain}, includes capabilities for
  %code and data segments as well as syscalls and signals.
  %
  \vspace{-6pt}
  \begin{defi}%[\subdomain]
    An \subdomain is a lightweight virtual machine represented by (instruction,
    \segment, entry-point, return-point, file, mapping, process, and exception)
    capabilities.
  \end{defi}
  \vspace{-6pt}
  %
  % --------------------
  Instruction capabilities specify which instructions are permitted without
  monitoring, and is required to fully virtualize the CPU---similar to the
  hosted architecture of VMMs, SFI, and Nested Kernel approaches.
  Explicitly representing instructions is critical as many protection models
  control instructions by using hardware privilege levels (rings), capability
  hardware, software based techniques like SFI (inline monitors), or
  deprivileging (static verifiers w/ runtime code integrity).
  %
  %As an example, recent work uses memory protection keys to isolate virtual
  %regions, however, the hardware exposes the key register to corruption through
  %\wrpkru.
  %
  %As we show in our prototype, we implement a restricted view of CPU state by
  %preventing any access to \wrpkru and \syscall instructions from non-\monitor
  %code, but do so using diverse mechanisms.
  %
  The way we virtualize the CPU also influences low-level mechanisms that
  enforce protected entry and exits.
  
  % --------------------
  Memory capabilities allow an \subdomain to read, write, or execute a
  \emph{\segment}, which is a subregion of the virtual address space.
  The default \segments for each \subdomain include: stack, heap, and code.
  File system capabilities specify operations permitted for opening, reading,
  and writing runtime state through the file system.
  Mapping capabilities determine what changes to the address space (\eg mmap,
  mprotect, \etc) a \subdomain has.
  Process capabilities determine what interactions are permitted between
  processes.
  Exception capabilities allow a \subdomain to securely register for and handle
  signals (\eg SIGSEGV). 
  Entry-point capabilities denote points at which a \subdomain transition is
  permitted, effectively converting function calls into RPCs.
  %and is much like converting function calls into an RPC for context-switching
  %and message passing.
  %
  Return-point capabilities are dynamically generated on cross-domain calls,
  \emph{\xcall}, and require the machine to return in nested order.
  % --------------------
  %Each \subdomain specifies a set of valid entry points from code \segment{}s
  %to preserve control integrity~\cite{Lampson:Protection:1974, dandvanhorn}.
  %
  Each \subdomain, by default, is granted exclusive access to its own code,
  data, and stack \segments.
  % --------------------
  An \emph{execution context} is the combination of the (\subdomain $X$ thread
  context).
  %, which includes the program counter, stack pointer, and other per domain CPU
  %registers.
  %
  Thus the \subdomain is similar to a traditional process by allowing multiple
  threads to coexist concurrently, while threads traverse \subdomains as they
  execute.
  %
  %This model allows for the greater range of flexibility for developing
  %extensible protection. 
  %
  %For the basic abstraction we choose not to tie the \subdomain abstraction to
  %per thread isolation, which can be implemented as an extension on top of
  %this, and is not necessary to achieve meaningful privilege separation as we
  %show in \boxing model.
  %
  %This execution model is the exact same as provided by Mondrix.
  %
  %The \monitor supports the following interface: program start, interrupted
  %state, signals, \moncall{}s, and \xcall{}s.

  %
  %This model explicitly leaves out methodology for sharing: by default a
  %\subdomain is declared with a set of capabilities for code, data, and stack
  %\segments, instruction caps, syscall caps, signal caps, and entry points.
  %
  \iffalse % don't think we need This does not preclude an extended interface
  that considers sharing, but our primary goal in this paper is to focus on the
  abstraction that allows us to most cleanly separate policy and mechanism for
  the purpose of developing a general purpose privilege virtualization
  framework, \ie one in which the privilege code runs side by side with less or
  alternatively privileged code (\eg Native Client, SVA, NK, ERIM, \etc).
  %
  \fi
  \vspace{-6pt}
  % --------------------
  \begin{prop}[\Subdomain Isolation]
    Each \subdomain is granted exclusive access to its code, data, and stack
    \segments, guaranteed secure entry/return, mapping capabilities for it's own
    \segments, and capabilities to OS level interfaces unless explicitly
    excluded for isolating other \subdomain state.
  \end{prop}
  \vspace{-6pt}
  % --------------------
  With these capabilities, the \visor exposes the ability to fully virtualize
  each resource while restricting access to privileged in-process state (\eg
  monitor memory).
  This is essential as many applications cannot be deployed without a certain
  level of access, but the monitor itself must ensure its protection.
  %
  %This is one of the most critical features gained under the \visor model
  %relative to existing ad hoc approaches.
  
  % --------------------
  %\input{domstruct}

%% file: org.tex
%%--------------------------------------------------------
%\subsection{The Manager}
\subsection{Nested \Monitor Organization}
\label{sec:principles}
%%--------------------------------------------------------

  % --------------------
  The \arch is a process model where a security monitor, the \emph{\monitor}, is
  nested within the address space with full authority.
  The \monitor multiplexes the process to enforce modularity in a set of
  \subdomains.
  The first goal of the \monitor is to self-isolate, \ie secure the \monitor
  state and \subdomain abstraction from \emph{\ud} bypass.
  %
  %This section explicitly details this architecture, leaving the protection
  %abstractions as extensions on top of this basic isolation.
  % --------------------
  %The objective of \arch is to split the process into two \subdomains: a nested
  %\td, the \monitor, and the \ud.

  % --------------------
  \paragraph{In-Process Policy}
  The \monitor is granted full authority to all process resources, and the \ud
  is granted access to all process resources except for the following: \monitor
  \segments; memory protection (\eg registers via \wrpkru) and direct OS call
  (\eg \syscall) instructions; file system operations that would allow access to
  \monitor \segments (\eg read/write /proc/self/mem/\monitor-\segment); address
  space manipulation (\eg \Code{mmap}) that would expose \monitor \segments; and
  signal capabilities that could otherwise use to bypass \segment isolation.
  %
  %In this way, the \monitor virtualizes privilege within the address space while
  %also inserting the \monitor in between the \ud and all privileged resources,
  %where the higher privilege state all protection state, including everything
  %that that could allow unmediated access by the lower privilege \ud. 
  %
  Just like any kernelized system, protected gates ensure the \monitor is
  securely entered into when a protection domain switch occurs.
  This architecture is similar to and inspired by the hosted VMM architecture
  and Nested Kernel Architecture.  
  \vspace{-5pt}
  \begin{defi}[\arch]
    An \arch is a split process model where the \monitor is nested within the
    address space.
  \end{defi}
  \vspace{-6pt}

  % --------------------
  The \monitor is responsible for exporting the basic \subdomain abstractions
  for all \ud \subdomains, thus enabling a new method for virtualizing
  subprocess resources and enforcing the following property:
  % --------------------
  \vspace{-6pt}
  \begin{prop}[Complete Mediation]
    A non-bypassable \monitor that is simple and guarantees isolation.
  \end{prop}
  \vspace{-6pt}
  % --------------------
  To achieve this the \visor enforces the following policies: secure loading and
  initialization so that all protection is configured appropriately;
  exports call gates for cross domain calls and ensures argument integrity and
  context-switching;
  inserts a monitor for all system calls so that they can be fully virtualized;
  monitors all address space and protection bit modifications to ensure
  isolation is not disabled;
  controls all signals so that they route through the \monitor before
  transitioning to any \ud \subdomain;
  and handles concurrency to support multi-threaded execution.

  % --------------------
  \paragraph{Interface}
  In the basic architecture, the \monitor transparently inserts itself and
  presents a minimal interface to the protected resources.
  All virtualized resources are accessed through calls into the \monitor.
  %
  %In the process model, this typically means only a system call interface as
  %that is the mechanism by which most resources are accessed and typically the
  %only resource that must passthrough the monitor.
  %
  Access to address spaces and file systems are monitored through the system
  call interface. 
  Other resources are memory based and since the \ud has no access to the
  \monitor state, there is no need for an explicit interface.
  %
  %We do not define an \subdomain creation/destruction interface as that is the
  %responsibility of the extension for implementing \subdomain modularity, which
  %we believe is best tailored to the application itself.
  %
  A \subdomain is created with an \monitor call that specifies a set of pages
  and entry points.
  
  % --------------------
  \input{sphere}

%% file: sphere.tex
%%--------------------------------------------------------
%\subsection{Distributed Flow Monitor}
%%--------------------------------------------------------
\paragraph{Protection Sphere and Transitive Capabilities}
%%--------------------------------------------------------

  % --------------------
  %The basic properties work while remaining within a single process, however,
  Most operating systems provide a \fork and \exec functionality for creating
  new processes.
  If left unconsidered, as all prior work does, then the \fork'ed process will
  be able to access privileged state.
  %
  %With the common threat model of a runtime compromise where an attacker forks,
  %would allow trivial bypass of the protections.
  %
  Furthermore, on \exec the runtime erases the process, which on first glance
  preserves the basic property, however, the new process could use the file
  system to access the parent's process state.
  In this work we choose to ensure 1) all forks include the same \monitor that
  enforces the policy in the new address space, and 2) retain system interface
  restrictions in an \exec'ed process. 

  % --------------------
  \vspace{-6pt}
  \begin{defi}[\Sphere]
  Each process defines an \emph{\sphere}, a context within which
  \segment isolation is guaranteed in and across all \fork{}ed address spaces. 
  \end{defi}
  \vspace{-6pt}
  %
  %A \sphere is a collection of processes where the virtual address space is
  %considered as one large security context, so that segment based isolation can
  %be maintained.
  %
  %Prior work neglected flows through system objects, which means an attacker
  %can \forks a process directly read memory or use system call layer to bypass
  %protections. 
  %
  The implication is that a \sphere provides a new realm of capability
  programming with the potential to tradeoff restricted runtimes for all forked
  processes (similar to work on Shill~\cite{Moore:Shill:2014} and left for
  future exploration).
  To ensure \segment isolation, the following property must be maintained.
  \vspace{-6pt}
  \begin{prop}[Fork+Exec Transivtivity]
  The \monitor transitively enforces the capabilities in each \fork{}ed process. 
  \end{prop}
  \vspace{-6pt}

%% file: design.tex
%%----------------------------------------------------------
\section{\system Design and Implementation}
\label{sec:design}
%%----------------------------------------------------------
  
  % --------------------
  \system is a userlevel only \visor system that fully virtualizes privilege and
  prevents bypass attacks.
  Beyond memory and CPU virtualization, it emphasizes full virtualization of
  system calls and signals, as well as exposes and addresses concurrency,
  multi-threaded, and multi-domain challenges. 
  \system injects the monitor into the application, as the \td \monitor, and
  removes the ability of the \ud to directly modify \emph{privileged state}.
  Privileged state includes: protection information (\pkru and memory mappings),
  code, \monitor code and data, direct system call invocation, raw signal
  handler data, CPU registers on transitions and control-flow, and system
  objects.
  The \monitor is inserted on startup by hooking all system call execution and
  initializing the protection state so that the \td is isolated with no files
  opened or mapped.

  \input{tm}
  \input{ivmemiso}
  \input{syscall}

  \input{sysflow}
  \input{sigvirt}
  \input{mt}
\input{md}

  \input{impl}

%% file: tm.tex
%%----------------------------------------------------------
\subsection{Threat Model and Assumptions}
%%----------------------------------------------------------

  %%--------------------------------------------------------
  %\review{add here the threats, focusing on the specific
  %ones for intra-process general}
  %% adding

  % --------------------
  %The primary objective of \system is to insert a nested system interface
  %monitor within a process address space to split the components in trusted and
  %untrusted domains.
  %
  We assume that an application is benign but potentially buggy: the same as
  prior work.
  We assume \system is free of memory corruption bugs and trust the OS and
  hardware implementation of protection keys.
  An attacker can use bugs to launch a buffer overflow that both deviates
  control and injects a payload: a shell script or return, jump, or data
  oriented program.
  We assume an attacker can leverage any instructions within the memory space in
  an attempt to launch a shell, fork processes, and in general exercise any
  system interface.
  The attacker can attempt confused deputy attacks on \system interfaces in
  order to launch TOCTOU and concurrency corruption.
  %
  %Our system call isolation will hold in the case of a malicious app, without
  %any intra-app security policy the attacker already has full access.
  %
  %Our isolation and virtualization will remain even if the attacker
  %successfully launches any type of code-reuse attack in the environment.
  %
  %\system enforces $W \oplus X$.
  % --------------------
  %\paragraph{Assumptions}
  %
  %We do not virtualize the MPK subsystem, thus eliminating its use by other
  %components in the process.
  %
  %
  %protections based upon trustworthy kernel interfaces and no ioctl extensions
  %that bypass: no-ioctl syscap. 
  %
  We do not consider architectural side-channels.

%% file: ivmemiso.tex
% --------------------
\subsection{Privilege and Memory Virtualization}
% --------------------

  % --------------------
  CPU and memory virtualization are the foundation of \visor isolation.
  Most of this work was generated by prior work in single-privilege level
  isolation (Nested Kernel, Erim, and Hodor). 
  We provide a concise description here, and place a more complete overview is
  provided in Appendix~\ref{sec:appmem}.
  \system uses MPK protection keys to assign \segments to \subdomains.
  In our prototype, all pages of the \td have key \Code{0} and all pages of the
  \ud have key \Code{2}.
  A \emph{virtual privilege switch} occurs when the \pkru is modified so that
  access is granted to all domains, which is secured by using \emph{instruction
  capabilities} to ensure the only \wrpkru instructions are inside secured call
  gates.
  To ensure these two are in place, we must ensure code, control, and memory
  integrity from system object, concurrency, and signal threats, which are
  detailed in subsequent sections.
  We add a new consideration for preventing processes from gaining control by
  switching into 32-bit compatibility mode, which changes how some instructions
  are decoded and executed.
  The security monitor code may not enforce the intended checks when executed in
  compatibility mode.
  Thus, we insert a short instruction sequence immediately after WRPKRU or
  XRSTOR instructions that will fault if the process is in compatibility mode.
  See Appendix~\ref{app:32bit-mode} for more details.

%% file: syscall.tex
% --------------------
\subsection{System Call Monitor and Handling}
\label{sec:syscallvirt}
% --------------------

%
%So we define the \emph{\nexpoline}, as a secure trampoline that sits in front
%of and ensures all access to instructions are guaranteed to operate.
%
%In our prototype we explore several different ways of realizing a \nexpoline,
%each with diverse security and performance tradeoffs (\eg randomization vs
%CET).

% --------------------
\system must ensure that access to system objects is virtualized.
We could place this monitor in the kernel, however, that would separate the
memory protection logic from the mechanism and create greater external
dependencies.
Furthermore, it would push the policy specification into the kernel, but the
abstractions supported need to be extensible and thus will endanger the whole OS.
Instead we observe that system resources are provided via \syscalls, whose semantics are stable and allow for
reasoning and enforcement of \subdomain isolation policies.
Additionally, \system will have greater portability if targeting POSIX.
Even if we locate the monitor in the kernel, we will have to add
extra context switches and layers of complexity in the kernel. % for handling
%the virtualization.

% --------------------
\system virtualizes system objects by monitoring control transfers
between the \ud and the OS through a novel in-address space monitor,
called the \emph{\nexpoline}.
%
%The \nexpoline, that
%
\begin{prop}[Nexpoline]
  All legitimate syscalls go through \monitor checks and virtualization.
\end{prop}
The basic ways \system does this include: 1) preventing all \syscall operations from \ud
\segments and 2) mediating and virtualizing all others.
%
%To ensure the \nexpoline, \system enforces two invariants: 1) no \syscalls are
%callable from \ud \segments and 2) all legitimate uses have complete mediation.
%
We could use a control-flow integrity monitor to provide both of these, like
CPI~\cite{Kuznetsov:CPI:2014}, but that would add unnecessary overhead, require compiler support, and violate our minimal mechanism principle.
Alternatively, we could extend the OS, but that would break our principle of
no kernel dependencies and cost.

% --------------------
\subsubsection{Passthrough}
The first step is to determine what virtualization, if
any, is necessary, because many \syscalls do not allow \subdomain bypass.
Additionally, a \syscall will use the application's virtual addresses to access the memory, which means that the
MPK domain will be enforced even from supervisor mode---this is
something we learned only through failing, so it is important to note that by
default the kernel leaves the MPK domain untouched and thus the hardware
continues to enforce the protection even from supervisor mode.
The benefit of this is that any kernel access to \subdomain \segments not
permitted by the current \pkru value will trap into the
\monitor---a powerful deny-by-default policy enforced even on
\Code{ioctls} with unknown semantics.
It does not mean that the kernel cannot remap pages and get around the domains, but a common path for access must be coded around, adding greater
confidence to the protected access paths.
%
%Virtualizing these \syscalls would require memory address checking which can be
%offloaded to the MPK hardware.
%
With these passthrough \syscalls, we use our protected \nexpoline control path
and right before executing the \syscall, we transition the \pkru domain to the
original caller so the kernel will respect the memory policies in place.
%
%The PKRU has been switched back so the kernel can treat the syscall as it has
%been called by the user directly and handle it properly.
%
%However, \system will still recognize the current state as \td and prevent data
%or control flow leaking when interrupted.
%
After the \syscall, \system will switch to \td to finalize the \syscall and
then transition back to the calling \subdomain.

% --------------------
\subsubsection{No \syscall from \ud \segments}
To prevent direct invocation of \syscalls, we remove all
\syscall instructions from \ud and ensure integrity like we do for \wrpkru,
however, the \syscall opcode is short and might lead to high false positives.
Instead, we use OS sandboxes that restrict \syscall to a protected \td
\segment.
There are two that can be used: seccomp and dispatch.
When we started, seccomp was the only option, but has many drawbacks:
1) you cannot grow or modify a seccomp filter, making support for multi-threading and forks challenging; and 2) it adds significant overhead.
The only way to address the second is to use a different mechanism.
Thus, we explore a recently released kernel dispatch mechanism,
a lightweight filter that restricts \syscalls to the particular \segment.
Both of these mechanisms work by specifying the virtual address region that is
permitted to invoke system calls, which we use to restrict to \monitor
\segments.

% --------------------
\subsubsection{Complete mediation for mapped \syscall}
Unfortunately, the only way to invoke a \syscall is for the opcode to exist in
the runtime, meaning it must be placed in the memory where the \ud can jump to.
Ideally, protection keys would distinguish executability and we could use a
\subdomain switch, but they do not: \intel relies on the NX mappings.
Alternatively, \segments with \syscall opcodes could be marked NX, but the
\nexpoline would require another \syscall to enable write access to the page.

% --------------------
%Instead, we define a call gate mechanism that ensures the \syscall instruction
%cannot be executed unless entering through the gate.
%
%Due to the low-level nature of the \nexpoline and diverse tradeoffs, we created
%several mechanisms.
%
Instead, the \nexpoline protects each instance of \sysretg instruction sequence,
i.e., the \srgadget, so that if control neglects to enter through the call gate,
the \syscall is inaccessible.
The basic control flow is to enter through the call gate and perform system
virtualization, set up the \nexpoline code \segment, jump to the \syscall, then
\Code{return} to the handler for cleanup.
%
%There are many ways to achieve at a single-privilege level and we explore three
%novel techniques.
%
%Isolation is provided by using mechanisms that to get access to the \syscall
%control-flow must enter through the entry gate, guaranteeing all \syscall
%virtualization policies are enforced.
%
We develop three isolation techniques: 1) randomizing the location of the
\srgadget and restrict access to the \monitor, 2) making the instruction ephemeral
by adding and removing it before and after each \syscall, and 3) using \intel CET
hardware.
These designs become complicated when considering concurrency, which we detail
in \S\ref{sec:concurrency}.

% --------------------
\paragraph{Randomized Location}
% --------------------
To abuse the \srgadget the attacker must know its location, which is randomized in the first isolation approach.
%
%As such, the first isolation approach randomizes the location of the \srgadget.
%
We create one pointer that points to the \srgadget, and make it readable by the
\monitor \subdomain.
This means that to get access to the pointer, the \subdomain must be switched-to
first, and thus guarantee protected entry.
The pointer is looked up immediately after switch, which means that all code
between that instruction and the \srgadget will execute: \monitor executes all
virtualization and once approved invokes the \srgadget.
This ensures complete mediation because the only way to get the \srgadget
location is to enter at the beginning. %, which ensures full virtualization.
The \srgadget can then be re-randomized at various intervals to provide stronger
or weaker security; we measure the cost of randomizing at differing numbers of
\syscalls.
\iffalse
%
Multi-threading creates some complexity as it could leak information, and we address this by
creating a per-thread pointer and giving enough virtual space to remain
probabilistic ally secure.
%
The benefit of this technique is that it is the simplest, and most of the time,
results in the best performance.
\fi

% --------------------
\paragraph{Ephemeral On-Demand}
% --------------------
While randomization---especially if randomizing on each \syscall---creates a
high degree of separation, it is not guaranteed.
To provide deterministic isolation, we present the \emph{ephemeral} \nexpoline,
which achieves isolation by writing the \srgadget into an executable \monitor
\segment on gate entry and rewriting to trap instructions ({\tt int 3}) after
completion.
This requires \system to create a single page for the \stramp in the \td with
read and write permission restricted to the \monitor (via MPK) and execute
permission for all domains.
\system ensures that when the \ud executes, the entire page is filled with {\tt
int 3} instructions, which would fault if the \ud were to jump to this
page.
%
%Upon entering the \td, \system updates the page with a single \srgadget opcode
%sequence.
%
%The \td then may perform system calls.
%
%Once the \td is ready to return to the \ud, it cleans up the previously written
%sequence by writing {\tt int 3} instructions.
%
The \monitor interposes on all control transfers from the OS to \ud, thus it
ensures that prior to any control transfer back to the \ud the \srgadget is
removed.
As a result, there is no executable \srgadget while
\ud is in control.

\iffalse
% --------------------
Handling multi-threaded execution is challenging because the \srgadget is
callable by other threads running in the process.
%
To address this issue, \system creates a per-thread filter that restricts each
thread's \syscall to only come from a per-thread \segment.
%
This means that the OS \syscall filter ensures that if a thread invokes the
\srgadget of another thread (while the system call is being handled) it will
trap.
%
This way, the \syscall instruction is ephemeral and only exists while the
thread is executing the \nexpoline.
%
This creates complexity as signals may modify the control flow of system calls,
which we describe in \S\ref{sec:sigvirt}.
\fi

% --------------------
\paragraph{Control enforcement technology (CET)~\cite{Intel:Control:2016}}
CET provides hardware to enforce control flow policies.
While designed for enforcing Control-Flow Integrity~\cite{Abadi:Control:2009},
we show how to (ab)use CET to implement a virtual call gate, which ensures
\sysretg is not directly executable by the \ud.
Briefly, CET guarantees that all returns return to the caller and indirect jumps
only target locations that are prefaced with the end-branch instruction.
CET also supports legacy code, by exporting a bitmap to mark all pages that can
bypass indirect jump enforcement, but the shadow stack must be used across the
whole application.

% --------------------
\system allocates a shadow stack for each \subdomain and ensures that a stack
cannot be used by a different \subdomain by assigning each one to a protected
\segment.
\iffalse
: for example, if the \ud attempts to change its stack or the \ud's
stack, the system will get a memory protection fault % due to the different key
from the currently executing \subdomain.
\fi
%
\system marks all \monitor entrypoints with \Code{ENDBR64}: denying transitions
into the \monitor from any indirect jumps.
This creates a problem though, because indirect jumps within the \monitor also
require end-branch instructions and could be used as alternative entrypoints to
the \monitor.
\iffalse
, but which would create entry points that might
bypass virtualization if called from the \ud.
\fi
%
Thus, all jumps within the \monitor are direct jumps with a fixed offset from
current IP and thus are not exploitable.
This allows \sysretg to be placed anywhere in the \td, since the hardware
automatically ensures all \syscall will start from a legit entrypoint.
While CET can provide greater security for the whole application, our evaluation
shows significant overheads compared to the other approaches (see \S\ref{sec:eval}).

%% file: sysflow.tex
% --------------------
\subsection{OS Object Virtualization} \label{sec:sysflow}
% --------------------

  % --------------------
  The primary goal of \system is to preserve \subdomain isolation, which
  requires system object virtualization for eliminating cross \subdomain flows.
  \system represents these in three three core system abstractions and policies
  to systematically reason about and specify policies: files (including
  sockets), address spaces, and processes.
  %
  %We first classify \syscalls into sensitive and non-sensitive.
  %
  %A sensitive system call is capable of breaking isolation, and non-sensitive
  %\system calls do not violate the isolation property and can use the
  %passthrough \nexpoline.
  %
  %The sensitive system calls break down into the following abstraction classes
  %along with associated policies.
  
  % --------------------
  \paragraph{Sensitive but Unvirtualized Syscalls} 
  A key class of system interfaces (\Code{ioctls}, \Code{sendto}, \etc) may
  index into regions of the address space that the kernel might accesses on
  behalf of a process, but as discussed, the kernel will use the userlevel
  virtual addresses which are protected by the hardware enforcing MPK domain
  isolation even from privilege accesses. 
  These do not require full system level virtualization, but if the kernel did
  not implement that strategy, they could be fully virtualized by analyzing the
  arguments and denying any access that crosses \subdomain isolation.
  %
  %A general issue across sensitive system calls are pointers or structures as
  %arguments.
  %
  %In order to ensure immutability in the presence of shared memory and
  %multi-processing, such arguments need to be deep copied before performing any
  %security policy evaluation.

  % --------------------
  \paragraph{Files} 
  The Linux kernel exposes (via the procfs) several sensitive files that may
  leak \subdomain's memory, because the kernel does not enforce page
  permissions, e.g., \Code{/proc/self/mems}.~\cite{Connor:PKU:2020}.
  To prevent any file-related system call from ever pointing to such a sensitive
  file, \system tracks the {\tt inode} of each opened file.
  Conveniently, {\tt inodes} are the same even when using soft or hard links.
  This allows \system to enforce that no open {\tt inode} matches the {\tt
  inode} of a sensitive file.
  The associated rules are transitively forwarded to child processes as they
  inherit the file descriptor table of the parent.

  % --------------------
  \paragraph{Mappings} 
  In addition, one may break the isolation property of \system by aliasing the
  same file mapping multiple times with different access permissions.
  For instance one mapping may allow read/execute, while the other alias mapping
  to the same file permits read/write accesses.
  We prevent such attacks by emulating the mapping using the regular file
  interface and copying the file to a read/write page first which is later
  turned read/execute after all security checks passed.
  As a result an executable page is never backed by a mapped file.
  
  % --------------------
  Memory system calls create, modify, or change access permissions of memory
  pages.
  Across such system calls we prevent \subdomain from accessing or altering
  another \subdomains memory, e.g. by never permitting a \subdomain to map
  another \subdomain's memory.
  In addition, new memory mappings by a \subdomain are tagged as belonging to
  the \subdomain.
  % --------------------
  \system enforces these policies by building a memory map that associates
  access permissions with \subdomain.
  
  \iffalse
  %
  %The state machine and transition functions of memory in
  %Figure~\ref{fig:attr_rel} summarize the polices.
  %
  %Basically, memory once marked for \domain $n$ cannot be marked for a
  %different domain.
  %
  %In addition, the figure depicts that memory isolation
  %techniques~\cite{Vahldiek-Oberwagner:ERIM:2019,Hedayati:Hodor:2019,
  %nacl,WebAssemblyCommunity:Security:2020} commonly require runtime validation
  %of any executable code protecting the technique's domain switch mechanisms or
  %bounds checks.

  A memory system call results in a state change of the memory map from its
  previous state.
  %
  When the memory system call is first intercepted, \system checks if the
  arguments target an existing memory mapping or create a new memory mapping.
  %
  With this information, the monitor protects the memory and ensures that the
  new state after performing the system call is allowed.
  %
  Once the system call is successful, we update the mapping table of the monitor
  to reflect the updated or new mapping.
  \fi

  % --------------------
  \paragraph{Processes}
  % --------------------
  \emph{[Remote Process Accessing]} 
  The Kernel permits virtual memory accesses via {\tt process\_vm\_readv} and
  {\tt process\_vm\_writev} system calls.
  These calls access memory of remote processes or the current process itself.
  For these two system calls, we apply the same restrictions as for file-backed
  system calls preventing a \domain from accessing another \domain's memory.
  In addition, we completely prevent access to another process' memory via {\tt
  process\_vm\_readv/writev}.
  % --------------------
  \emph{[fork and vfork]} Due to the insecure behavior of vfork, we emulate it
  by using {\tt fork} instead. {\tt fork} needs to be altered to enforce
  transitive policy enforcement across process boundaries.
  % --------------------
  \emph{[exec]} A process application can be modified using the {\tt exec}
  system call.
  In this case, the kernel loads the new executable and starts executing it.
  This is problematic, because we need to initialize its protections before the
  application.
  Hence, any {\tt exec} system call needs to be intercepted to ensure policy
  enforcement is enabled after {\tt exec}.
  \paragraph{Forbidden system calls} 
  Several \syscalls access protection state. 
  \system currently denies access to the following and leaves their
  virtualization to future work: {\tt clone} with shared memory, {\tt pkey\_*}
  system calls, {\tt modify\_ldt}, {\tt rt\_tgsigqueueinfo}, {\tt seccomp}, {\tt
  prctl} accessing seccomp, {\tt
  shmdt}, {\tt shmat}, {\tt ptrace}.

  % --------------------
  %\paragraph{Forking} can be used to create thread, and we simply banned that.

%% file: sigvirt.tex
  % --------------------
  \subsection{Signal virtualization}
  \label{sec:sigvirt}
  % --------------------
    % Nathan stupid simple list
    % 5) what about reentry while in IV cause bugs -> requires concurrency
    % control no reentry, 6) but no reentry breaks compat -> requires
    % deferring/sigpending, 7) what if signal delivered while in IV, can't
    % deliver right away because U need syscall to finish -> requires sign
    % pending and masking to simplify

    % --------------------
    Signals modify process control flow by pushing a signal frame onto the stack
    and transferring control to the point indicated by signal handler.
    Beyond exposing the \subdomains to control hijack, signals also expose the
    \pkru through \Code{struct sigframe}, which would allow an attacker to
    modify policies.
    \system virtualizes signals by adding a layer of indirection between \ud
    code and the OS.
    %
    %We describe the basic approach and detail corner cases.

    % --------------------
    \paragraph{Basic Operation}
    On signal registration, the \monitor tracks the handler in an internal
    table, \ie the virtual signal, and then registers a real signal with the OS
    that points to a handler within the \monitor.
    Just like interrupt handling in a kernel, \system splits signal handling
    into a top and bottom half.
    The bottom half receives a signal and sets up the top half to deliver it
    after performing a \Code{sigret} to the kernel.
    %
    %\system does this because 1) to avoid nested signals and 2) correctly
    %handle signals delivered asynchronously or synchronously while the \monitor
    %is executing.
    %
    \system only allows one signal to be queued for delivery at a time (by using
    signal masking), thus relying on the OS to manage nested signals without
    losing compatibility.
    %
    %Due to the atomicity needs of the \monitor, \system decouples the \ud
    %handler from the OS, by deferring the signal until after the OS pops it off
    %the signal stack.

    % --------------------
    \paragraph{Concurrency and Reentry}
    Signals can be delivered asynchronously---to the \ud or the \monitor---or
    synchronously to the \monitor---while the monitor was virtualizing a
    \syscall.
    %
    %This can cause bugs due to reentry, leading to security violations.
    %
    In each case, \system adds the signal to the pending queue and calls
    \Code{sigret}, however, the handler sets up different return locations based
    on the \subdomain.
    If it is the \ud, then \system modifies the return location to be the exit
    gate so that all configurations are in place for \ud execution.
    If in the \monitor, \system returns to the interrupted state preceding the
    signal. 
    This ensures that the monitor remains atomic and that all \system
    virtualization is cleaned up prior to returning to the \ud.
    %
    %To handle this \system creates an internal pending signals queue and
    %modifies the kernel's signal mask to defer signals delivered while in the
    %\monitor till the current task is completed: this allows one pending signal
    %in the \monitor and pushes the complexity of handling more to the OS.
    %
    %\system uses \Code{sigaction} to set the signal handler, it has a
    %\Code{sa\_mask} field in \Code{sigaction} structure, that we set to mask
    %signals when the kernel deliver signal to us.
    %
    %Once the current operation is completed, \system selects the last available
    %signal that has not been masked by the user and delivers it.

    % --------------------
    \paragraph{Default \pkru Domain}
    Linux resets the \pkru to a default value on all signal delivery, which
    means it can not deliver to the secure \monitor stack.
    We created a design and implementation to work with this limitation, but
    realized the signal handling interface should instead deliver the signal to
    the registering domain.
    Appendix~\ref{sec:appsig} describes this design as it is far more complex to
    securely handle signals, but in our final prototype we decided to modify the
    Linux kernel to deliver signals to the registering \pkru domain and thus
    securely transition from the kernel to the \monitor.

    % --------------------
    %\input{esyscall}
  
    \iffalse % TODO: this doesn't make sense but deadline and time to move
    \paragraph{CET} also complicated the design of signal by adding another
    stack that must take care of during the signal delivering. We a special
    syscall to write to the shadow stack whcih allows us to push RIP to restore
    address and RIP to signal handler and restore token on the shadow stack so
    we can have the required token for switching the stack when exiting \system.
    The similar trick is also used for virtualized sigreturn to switch to the
    old stack.
    %
    \fi

    \iffalse
    % we can detail this in the boxing section. It is a policy on top
    \paragraph{Multiple subdomains} As we discussed, control flow and
    corresponding CPU state are critical to the integrity of sensitive
    application. This applies to not only the \system but also the sandbox and
    the safebox. Since users can run whatever code in the subdomains, any
    interrpution during the execution of boxed code can be exploited to leak
    data. For this reason, we block the signal from the view of subdomains. The
    kernel can still deliver signal to \system signal entrypoint but we will
    treat it as a signal delivered in \td and pend that signal. 
    %
    \fi

%% file: mt.tex
% --------------------
\subsection{Multi-threading and Concurrency}
\label{sec:concurrency}
% --------------------
While concurrency is a well-known issue for multi-threaded monitors, prior
systems have ignored its design and implementation.
The issue is that the \subdomain abstraction allows concurrent threads and thus
exposes memory to corruption that could modify execution through TOCTOU attacks.

%-----------------
\paragraph{\Nexpoline Isolation with the Queen Thread}
An attacker could jump to the \srgadget when another thread is executing a
\syscall.
%
%Since we have 3 \nexpoline mechanisms, we need 3 different solutions.
%
The randomized \nexpoline is simplest since we depend on probability and can
enlarge the size of \srgadget region.
%and each thread takes separated \srgadget pages.
%
CET is also simple by using per-thread shadow stacks to prevent unauthorized
indirect jump.
The most complicated design is Ephemeral \nexpoline because it exposes the
\srgadget with executable permissions.
The solution is to create per thread \syscall filters that restrict each thread
to using a different page, so that even if another thread finds the location and
calls it, the kernel will deny.
Unfortunately, this design is complicated by the functionality of the filter.
The dispatch mechanism is not a problem since it can reset the region on any new
thread creation.
However, seccomp inherits filters in all forks.
\system addresses this by introducing a special thread, called Queen, that has
no seccomp filter and which spawns all new threads created by \Code{clone} with
a per thread \nexpoline filter.
%
%When \Code{clone} is called, \system requests Queen to handle it.
%
%Then, the Queen thread, which does not have a seccomp filter, clones a thread,
%applies a per-thread seccomp filter, and jumps to the start position of the new
%thread.

%-------------------
\paragraph{Monitor Atomicity}
TOCTOU attacks expose various state to corruption due to race conditions in
\monitor processing.
One solution is to use a monitor pattern that allows one thread to enter at a
time, however, this would be prohibitively costly.
Our solution is to use fine-grained locking based on the specific OS object
being interacted with.
First, we provide a lock for each file descriptor: each file descriptor can be
concurrently accessed, but only one thread can access one file descriptor in one
system call at a time.
Second, for mapping based OS objects, we maintain a global lock that only one
thread can call mapping based system calls.
Third, we provide one global lock for the system calls managing signals.

%% file: md.tex
\subsection{Multi-Domain for Least Privilege}
\label{sec:md}
% --------------------

So far we discussed the abstraction of a single untrusted \subdomain and a
privileged \monitor.
In this section we extend the abstraction to allow multiple untrusted
{\subdomain}es. This allows developers to decompose applications into
least-privilege functional units.
While \intel MPK is limited to 16 domains in total (and 14 usable for \system),
the design is independent of the number of domains and supports virtualizing of
domains as suggested by libMPK~\cite{Park:libmpk:2019}.
\system builds on MPK to provide an interface to isolate code and data, create a
\subdomain , switch {\subdomain}es via \xcall , and a utility library to assist
the needs of a \subdomain.

\paragraph{Secure Dynamic Creation of {\subdomain}es} We overload the existing
\monitor system call monitor to handle a virtual system call
\texttt{iv\_create\_domain(code, data, entrypoints, {\xcall}\_stub)}.
\system assigns an unused MPK domain to the \subdomain, and maps the code and
data to this domain to prevent other domains from accessing them.
It additionally virtualizes OS objects and signals as described in
\S~\ref{sec:sysflow} and \S~\ref{sec:sigvirt} and monitors the system
calls of all \subdomains to prevent privilege escalation.
Switching into the newly created \subdomain is restricted to the provided
entrypoints.
To improve the switching performance, \system installs a secure domain switch at
the \texttt{{\xcall}\_stub} which is used to call other \subdomains.

\paragraph{Securely Switching between \subdomains via {\xcall}} Switching
between \subdomains is supported via \texttt{\xcall(\subdomain, entrypoint\_id,
\ldots)}. Each switch is mediated via the \monitor which holds the information for
the entrypoints of each \subdomain. Therefore, \xcall switches to \monitor and
looks up the entrypoint of the target \subdomain without destroying the callee
arguments. It then updates the currently running \subdomain and switches to that
\subdomain (including a stack switch) to proceed to the function call. During
that process signal delivery to the current \subdomain is disabled to prevent
leakage via signals.

%% file: impl.tex
%%----------------------------------------------------------
\subsection{Implementation Details}
%%----------------------------------------------------------

%\todo{Update this section after finalizing all impl}

  % --------------------
  \system is built out of five primary components---secure
  loading, privilege and memory virtualization, \syscall
  virtualization, signal virtualization, and \xcall gates.
  We use the Graphene passthrough
  LibOS~\cite{Tsai:passthrulibos:2020,
  Tsai:Cooperation:2014, Tsai:GrapheneSGX:2017} to securely
  load, insert \syscall hooking into glibc, and separate the
  \td from \ud memory regions.
  We use ERIM~\cite{Vahldiek-Oberwagner:ERIM:2019} to
  isolate memory and protect \wrpkru, and 200 LoCs
  for tracking page attributes.
  We implement all \syscall and signal virtualization code.
  In total our system comprises $\approx$15k lines of
  code, with $\approx 6,400$ new \system code.

  \iffalse % NDD Cut for submittable draft bring back if needed
  % --------------------
  \subsection{Handling Non-GLIBC System Calls}
  % --------------------
    FANGFEI: There are risk involve in this Non-GLIBC syscall interface. I already give up this.
    This allows user to unmask certain signal inside signal handler. Allowing re-enter of signal entrypoint.

    % --------------------
    There are some libraries that execute system calls
    directly, for example, pthreads.
    %
    To deal with these we use a seccomp filter that creates
    a trap into \system.
    %

    \fi

%% file: usecase.tex
%%----------------------------------------------------------
\section{Programmable Least-Authority}
\label{sec:programmable}
%%----------------------------------------------------------

We demonstrate the extensibility of the \arch by developing virtual privilege
rings and using them to 1) eliminate the recent \Code{sudo}
vulnerability~\cite{CVE-2021-3156} and 2) \sandbox buggy parsers and isolate
sensitive OpenSSL data in NGINX.

% --------------------
%\input{privsep-bg}

% --------------------
\input{boxing}

% --------------------
\subsection{LibOS and Automated Transformation}
% --------------------
  % --------------------
  Programming the low-level interface affords power at the cost of effort
  required to use it.
  To simplify use we use a LibOS, called libsep, to aid developers in using the
  \sandbox and \safebox isolation, and we also provide an automation tool that
  separates at library boundaries.
  
  % --------------------
  \paragraph{libsep}
  % --------------------
  provides a simple interface for developers to statically annotate their code
  so that all data, code, and entry points are automatically separated during
  loading.
  Our approach uses section labels to do this.
  Additionally, the libsep tools automatically align \segments for encapsulation
  and generates the stub functions. 
  Instead of modifying the code with \xcall{}s, entrypoints are automatically
  wrapped to invoke the \xcall on use.
  Last, libsep provides a simple thread safe slab allocator.
  
  % --------------------
  \paragraph{Whole library isolation}
  % --------------------
  Alternatively \system can separate entire libraries.
  The developer adds a function that specifies the library virtual address and
  then the libos labels all exported symbols as entry points and generates stub
  code.
  A key element of this design is that domain transitions only occur in one
  direction because we use it as a \safebox.
  So if a library has a callback into a \sandbox or \unbox, that code will
  operate in the \safebox privilege level.

\subsection{Use Cases} 
\label{sec:usecase}
%--------------------------------

In this section we detail two use cases of \boxing.

%--------------------------------
\subsubsection{Eliminating \sudo Privilege Escalation} 
\label{sec:sudo_escal}
%--------------------------------
A recent bug was found in the \sudo argument parser that allows an attacker to
corrupt a function pointer and gain control with root access~\cite{CVE-2021-3156}.
We compartmentalize \sudo so that the parser code, in file \Code{parse\_args.c}, is
\sandbox{}ed, and restricted to only the command line arguments and an output
buffer.
The worst attack that can happen now is overflowing its internal buffer and eventually segfault or done nothing harmful.
In summary, by changing approximately 200 lines of codes, importing our \libos in \sudo and using \system,
we confine the argument parser and successfully prevent the root exploit.
More generally, almost all parsers have a similar type of behavior and could
benefit from similar changes, and possibly automatically.

%--------------------------------
\subsubsection{Towards a Least-Privilege NGINX} \label{sec:lp_nginx}
%--------------------------------
  % --------------------
  We present a novel compartmentalization of NGINX, which allows us to measure
  the effort required and performance costs of improving the security of a
  complex system while using the \boxing facilities.
  %
  %Additionally, we demonstrate the usefulness of our signal and \syscall
  %capabilities infrastructure.\ndd{not sure we do this though.}
  %
  %We make two observations: 1) parsers remain an entry point for buffer
  %overflows and full program compromise, and 2) there are many secrets in the
  %environment that any one of the \writeme{X nginx components could access and
  %leak}.
  %
  Our aim is to \sandbox the NGINX parser and \safebox the OpenSSL library.
  The value of the \sandbox is that it defends against common buggy entry points
  (\eg CVE-2009-2629, CVE-2013-2028, and CVE-2013-2070) used by attackers to
  launch control-flow hijacks (by injecting HTTP requests) and leaking or
  modifying sensitive state (\eg cookies, another client's responses, or keys).
  The value of the \safebox is to eliminate leakage of keys to the majority of
  the application, a TCB reduction argument, including session and private keys.

% --------------------
\paragraph{Safeboxing OpenSSL}
% --------------------
  %OpenSSL manages all the cryptographic keys for NGINX including the private
  %and session keys.
  %
  %Without safeboxing OpenSSL using \system, vulnerabilities in NGINX may leak
  %or modify the cryptographic keys of OpenSSL.
  %
  To prevent key leaks, we \safebox the entire OpenSSL libraries by using
  \libos, which links a special loader function that identifies all OpenSSL
  code and data sections during the start up time, and links \libos allocators,
  and initializes the \subdomain context.
  The tool also identifies entry points from the exports and instruments call
  gates for domain transitions.
  The developer minimally needs to link \libos into the application, identify
  the addresses of the sections, and call a new custom system call defined
  in \system, then \system does the rest---it can easily be applied to isolate
  other libraries.
  While this approach protects against vulnerabilities outside of OpenSSL, it
  does not prevent attacks from a compromised OpenSSL like
  Heartbleed~\cite{Durumeric:Matter:2014}.
  To prevent such attacks, we could split OpenSSL similar to  ERIM~\cite{Vahldiek-Oberwagner:ERIM:2019}.
  Another key element is that our abstraction only instruments calls into the
  library, so if the library calls back into the \unbox{}ed \domain, then it
  will be in the TCB.

  \iffalse
  %
  Prior efforts~\cite{:H2O:,Vahldiek-Oberwagner:ERIM:2019} isolate the keys in a
  separate process or memory domain.
  %
  While they achieve acceptable performance overhead ($<$2\%), such approaches
  reimplement large parts of the TLS module leading to new attack vectors and
  all, but ERIM~\cite{Vahldiek-Oberwagner:ERIM:2019}, do not protect
  frequently-accessed session keys.
  %
  \fi

% --------------------
\paragraph{Sandboxing HTTP Parser}
% --------------------
The Nginx parser is because its functionality is limited while it has been
historically buggy and grossly overprivileged.
While the parser only interacts with a small subset of data, that data is
referenced from a large structure, making sharing challenging.
We address this by allocating each instance of the structure into a special page
that the \sandbox is \grant{}ed permission to access upon invoking the parser
and permission is \revoke{}ed once finished.
This allows us to minimize overprivilege to only the per request data structure
and only for a short period of time, representing least-privilege.

%-----------------------------------
We manually modified NGINX HTTP request handler to identify the address of the
parser functions, and install the call gates by calling custom system call in
\system.
As well, we aligned the data structures with the page sizes to properly work
with MPK and we granted the data structures to the sandbox.

%% file: boxing.tex
%%--------------------------------------------------------
\subsection{Separation Facilities: Nested Boxing}
\label{sec:boxing}
%%--------------------------------------------------------

  % --------------------
  %Until now, \system guaranteed the integrity of the userlevel seperation
  %between the trusted and general untrusted domain by limiting the use of MPK
  %in untrusted code. However, this prevent any legit use of MPK to protect
  %private data and undermine the purpose of having MPK as a lightweight
  %multi-domain memory space. 
  %
  %Many prior works choose to use such a nested monitor as a place to put
  %sensitive code and data using an abstraction called the \emph{safe region}.
  %
  %For example, CPI puts metadata, ERIM/Hodor/..., 
  %%
  %Unfortunately, any code put into the safe region is now trusted and a part of
  %the security monitor. 
  %
  %Another issue with safe region is it keeps the best performance of protection
  %key by putting the domain switching into the original code, it leaves an
  %unclear boundary for the sensitive data, leaving which data should be
  %protected something based on the correct implementation of programmers. 
  %
  %Our next step is to further expose the ability of domain switching to
  %untrusted code with a limited semantics that prevent it from being abused. So
  %the user applications can fully utilize the native multi-domain feature from
  %CPU, while maintain the integrity of all other domains.

  % --------------------
  Least-authority is hard to apply in practice because security policies are
  highly dependent on the objects being protected.
  As indicated, many abstractions are rigid and do not allow for specialization
  from the application developers.
  The \arch allows us the ability to use the \monitor to explore diverse
  \subdomain and sharing models. 
  %
  %Many approaches can semi-automatically separate given a set of labels for
  %\sandbox code or \safebox data, however, these techniques neglect the
  %majority of potential opportunities and do not multiplex domain separation
  %within the nested environment. 
  %
  %To improve programmability and make use of the \nemo, 
  Thus, we present the \boxing abstraction, which effectively creates virtual
  privilege rings in the process.
  The \boxing model allows each level access to all resources of the less
  privileged layers, while removing the ability from those domains to access
  more privileged \domains. 
  In this paper we fix the number of domains to four from most to least
  privileged: \monitor, \safebox, \unbox, and \sandbox.
  Each \domain is given an initialized \subdomain that provides capabilities for
  accessing \domain resources.
  To make programming easier, we also use a libos that aids in allocation and
  separation policy management.

  \paragraph{Dynamic Memory Management}
  \label{sec:principles}
  %%--------------------------------------------------------
  % --------------------
  One of the core challenges with privilege separation is modifying the code so
  that data is statically and dynamically separated.
  Static separation is easily done using loader modification, but dynamic memory
  management is harder, in particular when we have to ensure \segment isolation.
  In our system we provide a \nemo allocator that transparently replaces
  whatever allocator the code originally used and automatically manages the heap
  and associated privilege policies.

  \iffalse
  You can still use amy memory allocator in the unboxed, but
  it is unsafe for both transfering the control to unprotected
  code with the privilege and also the memory allocated leave 
  in the unboxed memory. 
  
  %
  In general, the safebox should have its own allocator inside
  the safebox by using the syscalls to 
  map pages in the memory domain and providing the allocation interface,
  if it needs any dynamic memory. 
  This prevents unboxed code from hijacking the memory allocation
  which cases safebox to put sensitive data to unprotected memory. 
  %
  
  %
  To ease the design we provide a lightweight libos for
  managing the dynamic allocator that any box can include
  and use on it's memory regions.
  %
  \fi

  %%--------------------------------------------------------
  \paragraph{Memory Sharing}
  \label{sec:share}
  %%--------------------------------------------------------
  An \subdomain shares data through a simple manual page level \grant/\revoke
  model.
  A \subdomain \grant{}s access to any of its pages to a lower privilege \domain
  and removes access through the \revoke operation.

  %%--------------------------------------------------------
  \paragraph{Protected Entry and Return}
  \label{sec:xcall}
  %%--------------------------------------------------------
  % --------------------
  Cross \domain calls, or \xcall{}s, are invoked by the calling \domain and can
  only enter the called \domain at predefined entry points as specified by the
  \subdomain definition. 
  This interface will reject all attempts of accessing the \safebox if it is not
  to a preloaded entrypoint.  
  It will then do the \xswitch: switch the stack, current domain ID, store the
  return address in a protected memory \segment, and transfer control to the
  \safebox.
  When the called function finishes, it returns to the interface function, which
  \xswitch{}es back to the \ud. 
  Entry points can either be defined manually or as we show for full library
  separation, by using the library export list.
  This model of control flow allows the called domain to subsequently call less
  privileged code, \ie if it does this the called code operates within the
  \subdomain context and is thus in the TCB.
  We allow users to determine when and how to use these features, granting
  greater flexibility at the cost of more complexity in reasoning about security
  if a callback is issued.
  This can implement the Shreds abstraction, if used in code with no callbacks.

%% file: eval.tex
%%----------------------------------------------------------
\section{Evaluation}
\label{sec:eval}
%%----------------------------------------------------------

In this section we evaluate the security and performance of \system.
We highlight {\system}'s security properties protecting against
known~\cite{Connor:PKU:2020} and additional attacks found by us.
Subsequently, we investigate the performance characteristics of \system in
several microbenchmarks, on common applications, and least-privilege Nginx
use case.

\input{sec_eval}
\input{perf_eval}

\input{use_eval}

%% file: sec_eval.tex
%%----------------------------------------------------------
\subsection{Security Evaluation}
%%----------------------------------------------------------

\begin{table}
	\footnotesize
	\begin{tabular}{p{4.5cm}|c|c|c}
		\multicolumn{1}{c}{Attack} & \srand & \semph & CET \\ \hline \hline
		Inconsistency of PKU Permission~\cite{Connor:PKU:2020} &  $\bullet$ & $\bullet$ & $\bullet$\\ \hline
		Inconsistency of PT Permissions~\cite{Connor:PKU:2020} &  $\bullet$ & $\bullet$ & $\bullet$\\ \hline
		Mappings with Mutable Backings~\cite{Connor:PKU:2020} &  $\bullet$ & $\bullet$ & $\bullet$\\ \hline
		Changing Code by Relocation~\cite{Connor:PKU:2020} &  $\bullet$ & $\bullet$ & $\bullet$\\ \hline
		Modifying PKRU via sigreturn~\cite{Connor:PKU:2020} &  $\bullet$ & $\bullet$ & $\bullet$\\ \hline
		Race condition in Signal Delivery~\cite{Connor:PKU:2020} &  $\bullet$ & $\bullet$ & $\bullet$\\ \hline
		Race condition in Scanning~\cite{Connor:PKU:2020} &  $\bullet$ & $\bullet$ & $\bullet$\\ \hline
		Determination of Trusted Mappings~\cite{Connor:PKU:2020} &  $\bullet$ & $\bullet$ & $\bullet$\\ \hline
		Influencing Behavior with seccomp~\cite{Connor:PKU:2020} &  $\bullet$ & $\bullet$ & $\bullet$\\ \hline
		Modifying Trusted Mappings~\cite{Connor:PKU:2020} &  $\bullet$ & $\bullet$ & $\bullet$\\ \hline
		Forged Signal & $\bullet$ & $\bullet$  & $\bullet$\\ \hline
		Fork Bomb & $\circ$ & $\bullet$ & $\bullet$\\ \hline
		Syscall Arguments Abuse & $\bullet$ & $\bullet$ & $\bullet$\\ \hline
		TSX attack & $\circ$ & $\bullet$ & $\bullet$ \\\hline
        Race condition & $\bullet$ & $\bullet$ & $\bullet$
		\\ \hline
	\end{tabular}

	\caption{Quantitative security analysis based on attacks demonstrated
		in~\cite{Connor:PKU:2020} and attacks found by us. $\circ$ indicates the
		variant of \system in this column is vulnerable, $\bullet$ if it prevents
		this attack.
		%$\times$ indicates this attack is beyond \system's threat model.
        }
	\label{tab:qsa}
\end{table}

Table~\ref{tab:qsa} summarizes the quantitative security analysis based on known
attacks described by Conner \etal~\cite{Connor:PKU:2020} and additional attacks
we found.
\system defends against the attacks raised in~\cite{Connor:PKU:2020} and new
attacks we found by virtualizing and monitoring OS objects and preventing
privilege escalation via signals, multi-threading, multi-domain.
We briefly discuss new attack vectors and refer the reader for details to the
Appendix~\ref{sec:apdx_sec_eval}.
The attacks try to bypass \system by performing system calls modifying the
protection policies or try to elevate privileges by overriding the PKRU
register.

\paragraph{Forged Signal} The untrusted application may forge a signal frame
which includes the PKRU value and directly call the monitor’s signal entrypoint.

To differentiate forged signals from honest kernel signals, the signal
entrypoint in \system needs to be capable to distinguish both and stop forged
signals.

\paragraph{Fork Bomb} In case of a random \nexpoline, the random location can be
brute forced via a fork bomb that creates an infinite amount of processes trying
to make a system call by guessing the \nexpoline's location.
This bypasses the regular entrypoint of the \nexpoline and any security checks.
The ephemeral and CET \nexpoline are not vulnerable to this attack.

\paragraph{Syscall Argument Abuse} An attacker may use pointers in unvirtualized
system call that point to memory of a privileged domain to test for bits in the
contents.

\paragraph{TSX Attack} \intel TSX~\cite{intelTSX} supports transactional memory
which performs rollback of memory and execution.
This feature can be used to efficiently probe memory contents.
Only the random \nexpoline is vulnerable to this attack, since the secret random
location can be probed via TSX.

\paragraph{Race Conditions} Race conditions in \system could be exploited in
mutli-threaded and -processed applications.
Shared memory across multi-processes provides a potential avenue to modify
system call arguments by a second process after \system performed the necessary
security checks.
To prevent these multi-process attacks, \system relies on its own copy of the
system call arguments which cannot be access by another process.
Similar attacks may also be performed by multiple threads.
In particular, the interleaving of system calls can be exploited to gain access
to secrets.

%% file: perf_eval.tex
%%----------------------------------------------------------
\subsection{Performance}
%%----------------------------------------------------------

In this section we characterize the performance overhead of \system. First, we
explore microbenchmarks focussing on the cost to intercept system calls and
signals.
Second, we demonstrate the performance of \system for common applications.
Third, we evaluate the cost of the least-privilege Nginx use case.

%-----------------------------
We perform all experiments on an \intel 11th generation CPU (i7-1165G7) with
4 cores at 2.8GHz (Turboboost and hyper-threading disabled), 16GB memory
running Ubuntu 20.04, and the kernel version 5.9.8 with CET and syscall dispatch
support.
For all experiments we average over 100 repetitions and analyze different
\system\ configurations.
\system\ relies on a Seccomp filter or a syscall user dispatch (denoted by Sec
or Dis) for system call interception, and random, ephemeral, or CET trampoline
(denoted as rnd, emp, cet).
In this configuration space we evaluate 5 different configurations
($({secc|disp})\-({rand|eph|cet})$) excluding $disp-rand$ configuration because
it's major overhead is from the random number generation.
Throughout this section, we compare against MBOX~\cite{Kim:Practical:2013} and
strace, ptrace-based system call monitors.
MBOX fails for experiments using common applications.
In these cases we approximate the performance of MBOX using strace.
In our microbenchmarks strace outperforms MBOX by 2.7 \% providing a
conservative lower bound for MBOX.

%%%%%%%%%%%%%%%%%%%%%%%%%%%%%%%%
%% Shoot! I don't know why, but using macro in tikzpicture make the compilation freeze
%%%%%%%%%%%%%%%%%%%%%%%%%%%%%%%%
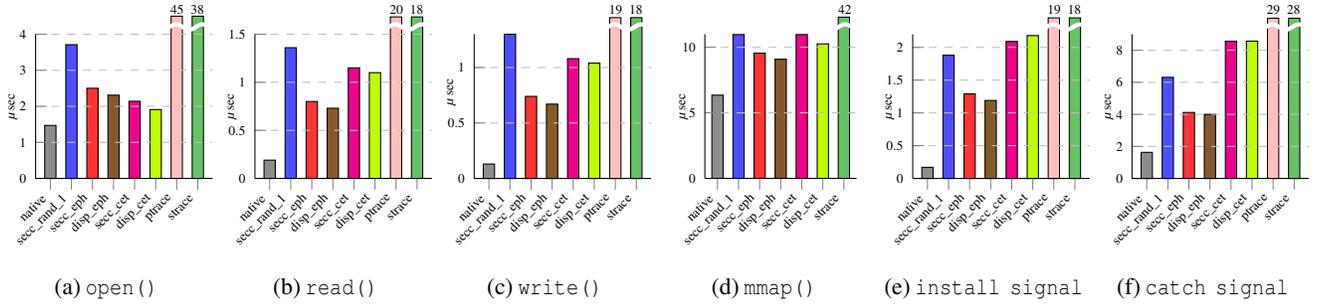
\begin{figure*}[t]
	\centering
	\begin{subfigure}{0.16\textwidth}
		\centering
		\noindent\begin{tikzpicture}
		\pgfplotsset{
			width=\textwidth,
			height=3.5cm,
			every tick label/.append style={font=\tiny},
		}
		\begin{axis}[
		ybar=-0.15cm,
		ylabel={$\mu$ sec},
		bar width=0.15cm,
		ymax=4,
		ymin=0,
		x=0.14cm,
		axis on top,
		restrict y to domain*=0:4.5,
		visualization depends on=rawy\as\rawy, % Save the unclipped values
		after end axis/.code={ % Draw line indicating break
			\draw [ultra thick, white, decoration={snake, amplitude=1pt}, decorate] (rel axis cs:0,1.05) -- (rel axis cs:1,1.05);
		},
		nodes near coords align={vertical},
		every node near coord/.append style={color=black},
		axis lines*=left,
		ymajorgrids=true,
		grid style=dashed,
		clip=false,
		ylabel style={at={(-0.05,0.5)}, font=\tiny},
		symbolic x coords={native,,secc\_rand\_1,,secc\_eph,,disp\_eph,,secc\_cet,,disp\_cet,,ptrace,,strace},
		x tick label style={rotate=45,anchor=east, font=\tiny},
		xtick={native,secc\_rand\_1,secc\_eph,disp\_eph,secc\_cet,disp\_cet,ptrace,strace},
		legend style={at={(0.45,-0.2)},
			anchor=north,legend columns=-1},
		]
		\addplot[fill=darkgray!60!white] coordinates {(native,1.47)};
		\addplot[fill=blue!70!white] coordinates {(secc\_rand\_1,3.71)};
		%\addplot[fill=magenta] coordinates{(secc\_rand\_16,2.14)};
		\addplot[fill=red!80!white] coordinates{(secc\_eph,2.5)};
		\addplot[fill=brown!70!black] coordinates{(disp\_eph,2.31)};
		\addplot[fill=magenta] coordinates{(secc\_cet,2.14)};
		\addplot[fill=lime] coordinates{(disp\_cet,1.91)};
		\addplot[fill=pink] coordinates{(ptrace,44.81)};
		\addplot[fill=darkgreen!60!white] coordinates {(strace,37.89)};
		\end{axis}
		\node at (1.88,2.25) {\tiny 45};
		\node at (2.15,2.25) {\tiny 38};
		\end{tikzpicture}
		\caption{\texttt{open()}}
		\label{fig:lmbench_open}
	\end{subfigure}
	\begin{subfigure}{0.16\textwidth}
		\centering
		\begin{tikzpicture}
		\pgfplotsset{
			width=\textwidth,
			height=3.5cm,
			every tick label/.append style={font=\tiny},
		}
		\begin{axis}[
		ybar=-0.15cm,
		ylabel={$\mu$ sec},
		bar width=0.15cm,
		ymax=1.5,
		ymin=0,
		x=0.14cm,
		axis on top,
		restrict y to domain*=0:1.68,
		visualization depends on=rawy\as\rawy, % Save the unclipped values
		after end axis/.code={ % Draw line indicating break
			\draw [ultra thick, white, decoration={snake, amplitude=1pt}, decorate] (rel axis cs:0,1.05) -- (rel axis cs:1,1.05);
		},
		nodes near coords align={vertical},
		every node near coord/.append style={color=black},
		axis lines*=left,
		ymajorgrids=true,
		grid style=dashed,
		clip=false,
		ylabel style={at={(-0.05,0.5)}, font=\tiny},
		symbolic x coords={native,,secc\_rand\_1,,secc\_eph,,disp\_eph,,secc\_cet,,disp\_cet,,ptrace,,strace},
		x tick label style={rotate=45,anchor=east, font=\tiny},
		xtick={native,secc\_rand\_1,secc\_eph,disp\_eph,secc\_cet,disp\_cet,ptrace,strace},
		legend style={at={(0.45,-0.2)},
			anchor=north,legend columns=-1},
		]
		\addplot[fill=darkgray!60!white] coordinates {(native,0.19)};
		\addplot[fill=blue!70!white] coordinates {(secc\_rand\_1,1.36)};
		%\addplot[fill=magenta] coordinates{(secc\_rand\_16,0.62)};
		\addplot[fill=red!80!white] coordinates{(secc\_eph,0.80)};
		\addplot[fill=brown!70!black] coordinates{(disp\_eph,0.73)};
		\addplot[fill=magenta] coordinates{(secc\_cet,1.15)};
		\addplot[fill=lime] coordinates{(disp\_cet,1.10)};
		\addplot[fill=pink] coordinates{(ptrace,19.60)};
		\addplot[fill=darkgreen!60!white] coordinates {(strace,18.05)};
		\end{axis}
		\node at (1.88,2.25) {\tiny 20};
		\node at (2.15,2.25) {\tiny 18};
		\end{tikzpicture}
		\caption{\texttt{read()}}
		\label{fig:lmbench_read}
	\end{subfigure}
	\begin{subfigure}{0.16\textwidth}
		\centering
		\begin{tikzpicture}
		\pgfplotsset{
			width=\textwidth,
			height=3.5cm,
			every tick label/.append style={font=\tiny},
		}
		\begin{axis}[
		ybar=-0.15cm,
		ylabel={$\mu$ sec},
		bar width=0.15cm,
		ymax=1.3,
		ymin=0,
		x=0.14cm,
		axis on top,
		restrict y to domain*=0:1.45,
		visualization depends on=rawy\as\rawy, % Save the unclipped values
		after end axis/.code={ % Draw line indicating break
			\draw [ultra thick, white, decoration={snake, amplitude=1pt}, decorate] (rel axis cs:0,1.05) -- (rel axis cs:1,1.05);
		},
		nodes near coords align={vertical},
		every node near coord/.append style={color=black},
		axis lines*=left,
		ymajorgrids=true,
		grid style=dashed,
		clip=false,
		ylabel style={at={(-0.05,0.6)}, font=\tiny},
		symbolic x coords={native,,secc\_rand\_1,,secc\_eph,,disp\_eph,,secc\_cet,,disp\_cet,,ptrace,,strace},
		x tick label style={rotate=45,anchor=east, font=\tiny},
		xtick={native,secc\_rand\_1,secc\_eph,disp\_eph,secc\_cet,disp\_cet,ptrace,strace},
		legend style={at={(0.45,-0.2)},
			anchor=north,legend columns=-1},
		]
		\addplot[fill=darkgray!60!white] coordinates {(native,0.13)};
		\addplot[fill=blue!70!white] coordinates {(secc\_rand\_1,1.30)};
		%\addplot[fill=magenta] coordinates{(secc\_rand\_16,0.56)};
		\addplot[fill=red!80!white] coordinates{(secc\_eph,0.74)};
		\addplot[fill=brown!70!black] coordinates{(disp\_eph,0.67)};
		\addplot[fill=magenta] coordinates{(secc\_cet,1.08)};
		\addplot[fill=lime] coordinates{(disp\_cet,1.04)};
		\addplot[fill=pink] coordinates{(ptrace,18.86)};
		\addplot[fill=darkgreen!60!white] coordinates {(strace,18.00)};
		\end{axis}
		\node at (1.88,2.25) {\tiny 19};
		\node at (2.15,2.25) {\tiny 18};
		\end{tikzpicture}
		\caption{\texttt{write()}}
		\label{fig:lmbench_write}
	\end{subfigure}
	\begin{subfigure}{0.16\textwidth}
		\centering
		\begin{tikzpicture}
		\pgfplotsset{
			width=\textwidth,
			height=3.5cm,
			every tick label/.append style={font=\tiny},
		}
		\begin{axis}[
		ybar=-0.15cm,
		ylabel={$\mu$ sec},
		bar width=0.15cm,
		ymax=11,
		ymin=0,
		x=0.14cm,
		axis on top,
		restrict y to domain*=0:12.3,
		visualization depends on=rawy\as\rawy, % Save the unclipped values
		after end axis/.code={ % Draw line indicating break
			\draw [ultra thick, white, decoration={snake, amplitude=1pt}, decorate] (rel axis cs:0,1.05) -- (rel axis cs:1,1.05);
		},
		nodes near coords align={vertical},
		every node near coord/.append style={color=black},
		axis lines*=left,
		ymajorgrids=true,
		grid style=dashed,
		clip=false,
		ylabel style={at={(-0.05,0.5)}, font=\tiny},
		symbolic x coords={native,,secc\_rand\_1,,secc\_eph,,disp\_eph,,secc\_cet,,disp\_cet,,strace},
		x tick label style={rotate=45,anchor=east, font=\tiny},
		xtick={native,secc\_rand\_1,secc\_eph,disp\_eph,secc\_cet,disp\_cet,strace},
		legend style={at={(0.45,-0.2)},
			anchor=north,legend columns=-1},
		]
		\addplot[fill=darkgray!60!white] coordinates {(native,6.35)};
		\addplot[fill=blue!70!white] coordinates {(secc\_rand\_1,10.97)};
		%\addplot[fill=magenta] coordinates{(secc\_rand\_16,9.24)};
		\addplot[fill=red!80!white] coordinates{(secc\_eph,9.55)};
		\addplot[fill=brown!70!black] coordinates{(disp\_eph,9.09)};
		\addplot[fill=magenta] coordinates{(secc\_cet,10.97)};
		\addplot[fill=lime] coordinates{(disp\_cet,10.26)};
		\addplot[fill=darkgreen!60!white] coordinates {(strace,42.38)};
		\end{axis}
		\node at (1.86,2.25) {\tiny 42};
		\end{tikzpicture}
		\caption{\texttt{mmap()}}
		\label{fig:lmbench_mmap}
	\end{subfigure}
	\begin{subfigure}{0.16\textwidth}
		\centering
		\begin{tikzpicture}
		\pgfplotsset{
			width=\textwidth,
			height=3.5cm,
			every tick label/.append style={font=\tiny},
		}
		\begin{axis}[
		ybar=-0.15cm,
		ylabel={$\mu$ sec},
		bar width=0.15cm,
		ymax=2.2,
		ymin=0,
		x=0.14cm,
		axis on top,
		restrict y to domain*=0:2.45,
		visualization depends on=rawy\as\rawy, % Save the unclipped values
		after end axis/.code={ % Draw line indicating break
			\draw [ultra thick, white, decoration={snake, amplitude=1pt}, decorate] (rel axis cs:0,1.05) -- (rel axis cs:1,1.05);
		},
		nodes near coords align={vertical},
		every node near coord/.append style={color=black},
		axis lines*=left,
		ymajorgrids=true,
		grid style=dashed,
		clip=false,
		ylabel style={at={(-0.05,0.5)}, font=\tiny},
		symbolic x coords={native,,secc\_rand\_1,,secc\_eph,,disp\_eph,,secc\_cet,,disp\_cet,,ptrace,,strace},
		x tick label style={rotate=45,anchor=east, font=\tiny},
		xtick={native,secc\_rand\_1,secc\_eph,disp\_eph,secc\_cet,disp\_cet,ptrace,strace},
		legend style={at={(0.45,-0.2)},
			anchor=north,legend columns=-1},
		]
		\addplot[fill=darkgray!60!white] coordinates {(native,0.17)};
		\addplot[fill=blue!70!white] coordinates {(secc\_rand\_1,1.88)};
		%\addplot[fill=magenta] coordinates{(secc\_rand\_16,1.13)};
		\addplot[fill=red!80!white] coordinates{(secc\_eph,1.29)};
		\addplot[fill=brown!70!black] coordinates{(disp\_eph,1.19)};
		\addplot[fill=magenta] coordinates{(secc\_cet,2.09)};
		\addplot[fill=lime] coordinates{(disp\_cet,2.18)};
		\addplot[fill=pink] coordinates{(ptrace,19.48)};
		\addplot[fill=darkgreen!60!white] coordinates {(strace,17.99)};
		\end{axis}
		\node at (1.88,2.25) {\tiny 19};
		\node at (2.15,2.25) {\tiny 18};
		\end{tikzpicture}
		\caption{\texttt{install signal}}
		\label{fig:lmbench_signal_install}
	\end{subfigure}
	\begin{subfigure}{0.16\textwidth}
		\centering
		\begin{tikzpicture}
		\pgfplotsset{
			width=\textwidth,
			height=3.5cm,
			every tick label/.append style={font=\tiny},
		}
		\begin{axis}[
		ybar=-0.15cm,
		ylabel={$\mu$ sec},
		bar width=0.15cm,
		ymax=9,
		ymin=0,
		x=0.14cm,
		axis on top,
		restrict y to domain*=0:10,
		visualization depends on=rawy\as\rawy, % Save the unclipped values
		after end axis/.code={ % Draw line indicating break
			\draw [ultra thick, white, decoration={snake, amplitude=1pt}, decorate] (rel axis cs:0,1.05) -- (rel axis cs:1,1.05);
		},
		nodes near coords align={vertical},
		every node near coord/.append style={color=black},
		axis lines*=left,
		ymajorgrids=true,
		grid style=dashed,
		clip=false,
		ylabel style={at={(-0.05,0.5)}, font=\tiny},
		symbolic x coords={native,,secc\_rand\_1,,secc\_eph,,disp\_eph,,secc\_cet,,disp\_cet,,ptrace,,strace},
		x tick label style={rotate=45,anchor=east, font=\tiny},
		xtick={native,secc\_rand\_1,secc\_eph,disp\_eph,secc\_cet,disp\_cet,ptrace,strace},
		legend style={at={(0.45,-0.2)},
			anchor=north,legend columns=-1},
		]
		\addplot[fill=darkgray!60!white] coordinates {(native,1.63)};
		\addplot[fill=blue!70!white] coordinates {(secc\_rand\_1,6.31)};
		%\addplot[fill=magenta] coordinates{(secc\_rand\_16,4.1)};
		\addplot[fill=red!80!white] coordinates{(secc\_eph,4.12)};
		\addplot[fill=brown!70!black] coordinates{(disp\_eph,4.00)};
		\addplot[fill=magenta] coordinates{(secc\_cet,8.56)};
		\addplot[fill=lime] coordinates{(disp\_cet,8.57)};
		\addplot[fill=pink] coordinates{(ptrace,29.36)};
		\addplot[fill=darkgreen!60!white] coordinates {(strace,28.08)};
		\end{axis}
		\node at (1.88,2.25) {\tiny 29};
		\node at (2.15,2.25) {\tiny 28};
		\end{tikzpicture}
		\caption{\texttt{catch signal}}
		\label{fig:lmbench_signal_catch}
	\end{subfigure}
	\caption{\label{fig:lmbench} System call latency of LMBench benchmark.
		Std. dev. below 5.7\%.}
	% All the data is measured on 01/17/2021
\end{figure*}

%-------------------------------
\subsubsection{Microbenchmarks}
%-------------------------------

%---------------------------------------------
\paragraph{System call overhead}
%---------------------------------------------
We evaluate \system's overhead on system calls and signal delivery in comparison
to native and the ptrace-based techniques.
Figure \ref{fig:lmbench} depicts the latency of LMBench
v2.5~\cite{McVoy:lmbench:} for common system calls.
Each~\system configuration and the ptrace-based techniques intercept system
calls and provide a virtualized environment to LMBench while protecting its
privileged state.

%----------------------------
\semph and \srand\_1 modify the trampoline on every system call, but \semph
saves the cost of randomizing the trampoline location and hence, incurs less
overhead.
\semph/\sdemph, and \sscet/\sdcet demonstrate the performance difference between
using a Seccomp filter or syscall user dispatch to intercept system call
invocations.
%
%----------------------------------
Overall, \sdemph outperforms all other configurations, while \srand\_1 is the
slowest.
Even though CET relies on hardware support, it does not outperform other
configurations.
\system adds 0.5 - 2 usec per system call for \sdemph for policy enforcement and
domain switches.
In comparison the ptrace-based technique incurs about 20 usec per invocation
which is 4.7-26.8 times slower than \sdemph.
%
%----------------------------------
We observe high overheads for \system protecting fast system calls like
\texttt{read} or \texttt{write} 1 byte (126\%-900\%), whereas long lasting
system calls like \texttt{open} or \texttt{mmap} only observe 29\%-150\%
overhead.
%
%---------------------------
\begin{figure}[t]
	\begin{tikzpicture}
	\pgfplotsset{
		width=6.8cm,
		height=4.3cm,
		every tick label/.append style={font=\footnotesize},
	}
	\begin{axis}[
	xlabel={read size [KB]},
	xlabel style={font=\footnotesize},
	ylabel={Normalized throughput},
	ylabel style={at={((-0.08,0.5)}, font=\footnotesize},
	xmin=1, xmax=512,
	log ticks with fixed point,
	ymin=0, ymax=1,
	x tick label style={/pgf/number format/1000 sep=\,},
	xtick={data},
	%x tick label style={font=\scriptsize},
	y tick label style={font=\footnotesize},
	xmode=log,
	log basis x={2},
	legend pos=south east,
	legend style={at={(1.25,0.9),font=\scriptsize},
		anchor=north,legend columns=1},
	ymajorgrids=true,
	grid style=dashed,
	]
	\addplot[color=blue!70!white, mark=square]
	coordinates { (1,0.491)(2,0.567)(4,0.661)(8,0.774)(16,0.856)(32,0.919)(64,0.958)(128,0.979)(256,0.993)(512,0.999)};
	\addlegendentry{\srand\_16}
	\addplot[color=red, mark=star]
	coordinates { (1,0.420)(2,0.493)(4,0.594)(8,0.723)(16,0.819)(32,0.895)(64,0.943)(128,0.971)(256,0.987)(512,0.996)};
	\addlegendentry{\semph}
	\addplot[color=brown!70!black, mark=diamond]
	coordinates { (1,0.455)(2,0.528)(4,0.629)(8,0.753)(16,0.840)(32,0.909)(64,0.950)(128,0.974)(256,0.990)(512,0.999)};
	\addlegendentry{\sdemph}
	\addplot[color=magenta, mark=triangle]
	coordinates { (1,0.304)(2,0.368)(4,0.470)(8,0.607)(16,0.731)(32,0.843)(64,0.913)(128,0.954)(256,0.979)(512,0.993)};
	\addlegendentry{\sscet}
	\addplot[color=lime, mark=+]
	coordinates { (1,0.315)(2,0.381)(4,0.484)(8,0.621)(16,0.745)(32,0.852)(64,0.916)(128,0.955)(256,0.978)(512,0.990)};
	\addlegendentry{\sdcet}
	\addplot[color=pink!85!black, mark=o]
	coordinates { (1,0.023)(2,0.031)(4,0.047)(8,0.079)(16,0.136)(32,0.232)(64,0.369)(128,0.535)(256,0.696)(512,0.818)};
	\addlegendentry{ptrace}

	\end{axis}

	\end{tikzpicture}
	\caption{\label{fig:fileio} Normalized latency of reading a 40MB file.
		Std. dev. below 1.5\%.}
	% All the data is measured on 01/17/2021
\end{figure}
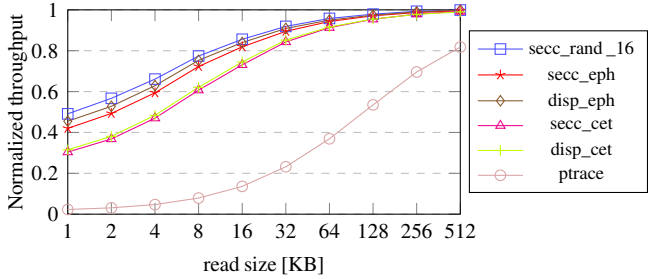
We demonstrate the difference by performing a throughput file IO experiment.
Figure \ref{fig:fileio} shows high overheads for reading small buffer sizes
which amortize with larger buffer sizes.
Since overhead induced by \system is per \syscall\ basis, to read a file with
bigger buffer size has much less overhead than with the smaller buffer size.
Even though we observe high overheads for some system calls, applications
infrequently use them and observe far less overhead as shown for common
applications in \S~\ref{sec:overhead_apps}.
%
%--------------------------------------------------
\begin{figure}[t]
	\begin{tikzpicture}
	\pgfplotsset{
		width=\textwidth,
		height=3.5cm,
		every tick label/.append style={font=\scriptsize},
	}
	\begin{axis}[
	%small,
	ybar=-0.3cm,
	ylabel={$\mu$ sec},
	bar width=0.3cm,
	ymax=1,
	ymin=0,
	x=0.3cm,
	axis on top,
	ytick={0.2,0.4,0.6,0.8,1.0},
	ymajorgrids=true,
	grid style=dashed,
	nodes near coords align={vertical},
	every node near coord/.append style={font=\scriptsize,color=black},
	axis lines*=left,
	clip=false,
	ylabel style={at={(-0.07,0.5)}, font=\scriptsize},
	symbolic x coords={native,,secc\_rand\_1,,secc\_rand\_2,,secc\_rand\_4,,secc\_rand\_8,,secc\_rand\_16,,secc\_rand\_32,,secc\_rand\_1024,,secc\_eph},
	x tick label style={rotate=30,anchor=east},
	xtick={native,secc\_rand\_1,secc\_rand\_2,secc\_rand\_4,secc\_rand\_8,secc\_rand\_16,secc\_rand\_32,secc\_rand\_1024,secc\_eph},	%Don't know why but compile hangs when use macro
	]
	\addplot[fill=darkgray!60!white] coordinates {(native,0.1)};
	\addplot[fill=blue!70!white] coordinates {(secc\_rand\_1,1.05)};
	\addplot[fill=blue!65!white] coordinates {(secc\_rand\_2,0.67)};
	\addplot[fill=blue!60!white] coordinates {(secc\_rand\_4,0.45)};
	\addplot[fill=blue!55!white] coordinates {(secc\_rand\_8,0.35)};
	\addplot[fill=blue!50!white] coordinates {(secc\_rand\_16,0.29)};
	\addplot[fill=blue!45!white] coordinates {(secc\_rand\_32,0.27)};
	\addplot[fill=blue!40!white] coordinates {(secc\_rand\_1024,0.24)};
	\addplot[fill=red!80!white] coordinates {(secc\_eph,0.49)};
	\end{axis}
	\end{tikzpicture}
	\caption{\label{fig:rand} latency for \texttt{getppid} for different rerandomization scaling.
		Std. dev. below 6.7\%.}
	% All the data is measured on 01/17/2021
\end{figure}
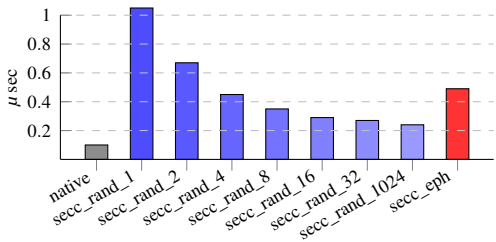

\paragraph{Randomization and performance tradeoff}
The \srand configuration rerandomizes the trampoline for each system call
generating an random number using \texttt{RDRAND} (approx. 460 cycles).
We explore alternative rerandomization frequencies to amortize the cost of
randomizing over several system calls.
We tradeoff performance with security, since the system call address is
simpler to guess if rerandomization happens less frequently.
The goal is to find a reasonably secure, but fast rerandomization frequency.
%
%---------------------------------
Figure \ref{fig:rand} evaluates \texttt{getppid} system call for different
randomization frequencies.
\texttt{getppid} is the fastest system call and hence, results in the highest
overhead of \system.
The overhead of \srand\ amortizes with less frequent randomization and does
not improve much beyond 16 system calls per randomization.
\srand\ at 4 system calls per randomization shows similar performance with
 \semph's performance which we also observed for other LMBench microbenchmarks.

%---------------------------
\begin{figure}[t]
	\begin{tikzpicture}
	\pgfplotsset{
		width=6.8cm,
		height=4.3cm,
		every tick label/.append style={font=\footnotesize},
	}
	\begin{axis}[
	xlabel={Number of threads},
	xlabel style={at={(0.5,-0.09)}, font=\footnotesize},
	ylabel={Bandwidth [GB/s]},
	ylabel style={at={(-0.06,0.5)}, font=\footnotesize},
	xmin=1, xmax=32,
	log ticks with fixed point,
	ymin=0, ymax=25,
	x tick label style={/pgf/number format/1000 sep=\,},
	%x tick label style={font=\scriptsize},
	y tick label style={font=\footnotesize},
	ytick={5,10,15,20,25},
	xmode=log,
	log basis x={2},
	legend pos=south east,
	legend style={at={(1.25,0.95),font=\scriptsize},
		anchor=north,legend columns=1},
	ymajorgrids=true,
	grid style=dashed,
	]
	\addplot[color=darkgray, mark=x]
	coordinates { (1,6.66236)(2,13.20222)(4,23.00738)(8,23.03954)(16,22.96551)(32,23.04999)};
	\addlegendentry{native}
	\addplot[color=blue!70!white, mark=square]
	coordinates { (1,5.53561)(2,10.96192)(4,19.20955)(8,19.14877)(16,18.7406)(32,17.69576)};
	\addlegendentry{\srand\_16}
	\addplot[color=red, mark=star]
	coordinates { (1,5.21335)(2,10.18612)(4,17.07920)(8,16.97985)(16,16.84076)(32,16.55245)};
	\addlegendentry{\semph}
	\addplot[color=brown!70!black, mark=diamond]
	coordinates { (1,5.44920)(2,10.90362)(4,19.04357)(8,18.98738)(16,18.5769)(32,17.5894)};
	\addlegendentry{\sdemph}
	\addplot[color=magenta, mark=triangle]
	coordinates { (1,4.47542)(2,8.60584)(4,15.57608)(8,14.13077)(16,12.16447)(32,9.38844)};
	\addlegendentry{\sscet}
	\addplot[color=lime, mark=+]
	coordinates { (1,4.61992)(2,8.83707)(4,15.88945)(8,14.41179)(16,12.38043)(32,9.51683)};
	\addlegendentry{\sdcet}
	\addplot[color=darkgreen!60!white, mark=o]
	coordinates { (1,0.755)(2,1.365)(4,2.311)(8,2.575)(16,2.492)(32,2.183)};
	\addlegendentry{strace}
	\end{axis}

	\end{tikzpicture}
	\caption{\label{fig:sysbench} Random read bandwidth for diff. number of
		threads measured with \texttt{sysbench}.
		Std. dev. below 0.7\%.}
	% All the data is measured on 01/17/2021
\end{figure}
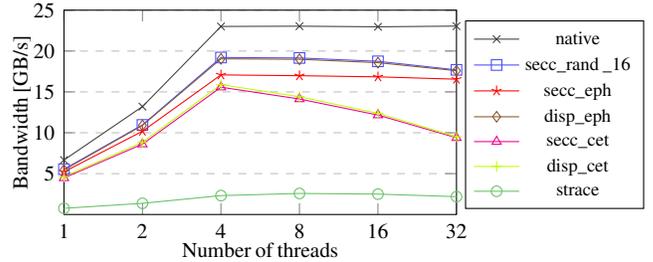

%%%%%%%%%%%%%%%%%%%%%%%%%%%%%%%%%
\paragraph{Thread scalability}
To prevent race conditions and TOCTOU attacks in \system, locks protect
\system's policy enforcement as addressed in the \S~\ref{sec:concurrency}.
We demonstrate the scalability of \system in figure~\ref{fig:sysbench} using the
sysbench~\cite{:sysbench:} tool which concurrently reads a 1 GB file from
varying number of threads.
Due to the additional locks in \system, the number of \texttt{futex} system
calls increases with the number of threads.

%----------------------------------------
At 4 threads all CPU cores are busy and we observe the best performance.
The overhead of each configuration is similar to the microbenchmarks.
\sscet and \sdcet suffer a performance decrease of up to 60\%, because the
syscall performance of CET-based configurations is the lowest.
Compared to strace, \system outperforms by 4.3-8.2 times.

%-----------------------------------------------
\subsubsection{Overhead on Applications}
\label{sec:overhead_apps}

%-------------------------------
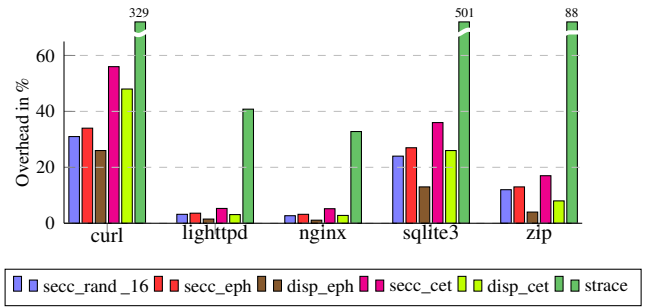
\begin{figure}[t]
	\begin{tikzpicture}
	\pgfplotsset{
		width=\textwidth,
		height=4cm,
		every tick label/.append style={font=\scriptsize},
	}
	\begin{axis}[
	ybar=1pt,
	enlarge x limits=0.1,
	ylabel={Overhead in \%},
	ylabel style={font=\scriptsize},
	bar width=4pt,
	ymax=65,
	ymin=0,
	restrict y to domain*=0:72,
	visualization depends on=rawy\as\rawy,
	after end axis/.code={
		\draw [ultra thick, white, decoration={snake, amplitude=1pt}, decorate] (rel axis cs:0,1.05) -- (rel axis cs:1,1.05);
	},
	axis on top,
	nodes near coords align={vertical},
	every node near coord/.append style={font=\scriptsize},
	axis lines*=left,
	clip=false,
	ymajorgrids=true,
	grid style=dashed,
	ylabel style={at={((-0.05,0.5)}},
	symbolic x coords={curl,lighttpd,nginx,sqlite3,zip},
	x tick label style={anchor=center,font=\footnotesize},
	xtick=data,
	legend style={at={(0.5,-0.25),font=\scriptsize},
		anchor=north,legend columns=-1},
	]
	\addplot [fill=blue!50!white] coordinates {(curl,31) (lighttpd,3.2) (nginx,2.7) (sqlite3,24) (zip,12)};
	\addplot [fill=red!80!white] coordinates {(curl,34) (lighttpd,3.6) (nginx,3.2) (sqlite3,27) (zip,13)};
	\addplot [fill=brown!70!black] coordinates {(curl,26) (lighttpd,1.5) (nginx,1.1) (sqlite3,13) (zip,4)};
	\addplot [fill=magenta] coordinates {(curl,56) (lighttpd,5.3) (nginx,5.2) (sqlite3,36) (zip,17)};
	\addplot [fill=lime] coordinates {(curl,48) (lighttpd,3.1) (nginx,2.8) (sqlite3,26) (zip,8)};
	\addplot [fill=darkgreen!60!white] coordinates {(curl,329) (lighttpd,40.8) (nginx,32.8) (sqlite3,501) (zip,88)};
	\legend{\srand\_16,\semph,\sdemph,\sscet,\sdcet,strace}
	\end{axis}
	\node at (1,2.8) {\tiny 329};
	\node at (5.33,2.8) {\tiny 501};
	\node at (6.75,2.8) {\tiny 88};

	\end{tikzpicture}
	\caption{\label{fig:apps_perf} Normalized overhead of diff. Linux applications.
		Std. dev. below 2.4\%.}
	% All the data is measured on 01/17/2021
\end{figure}

Along with the microbenchmarks, we analyze the performance of common
applications such as lighttpd~\cite{:Lighttpd:}, Nginx~\cite{:NGINX:},
curl~\cite{:CURL:}, SQLite database~\cite{:SQLite:}, and zip~\cite{:zip:}
protected by \system.
Figure~\ref{fig:apps_perf} shows the overall overhead of each application
compared to the native execution`.

%----------------------------
\paragraph{\texttt{curl}~\cite{:CURL:}} downloads a 1 GB file from a local web server.
It is particularly challenging workload for \system, since \texttt{curl} makes a
system call for every 8 KB and frequently installs signal handlers.
In total it calls more than $130,000$ \Code{write} system calls and more than
$30,000$ \Code{rt\_sigaction()} system calls to download a 1 GB file.
However, libcurl supports an option not to use signal, which reduces the
overhead about 10\% in average for \system but strace gets worse about 140\%.

%-----------------------
\paragraph{Lighttpd~\cite{:Lighttpd:} and Nginx~\cite{:NGINX:}} serve a 64 KB
file requested 1,000 times by an apachebench tool~\cite{:Apache:} client on the
same machine.
All configurations perform within 94\% of native.
\sdemph outperforms all other configurations and highlights \system's ability to
protect applications at near-zero cost with a throughput degradation of 1\%.
In contrast, strace has about 30\% overhead.

%---------------------------------
\paragraph{SQLite~\cite{:SQLite:}} runs its speedtest
benchmark~\cite{:SQLite:} and performs \texttt{read()} and \texttt{write()}
system calls with very small buffer size to serve individual SQL requests.
Contrary to the microbenchmarks, difference between configurations is larger.
Configurations using syscall user dispatch (\sdemph and \sdcet) observe about
30\% less overhead when compared to their Seccomp alternatives (\semph and
\sscet).
Strace performs poorly at more than 500\% overhead.

\paragraph{zip~\cite{:zip:}} compresses the full Linux kernel 5.9.8 source tree,
a massive task which opens all files in the source tree, reads their contents,
compresses them, and archives them into a zip file.
The observed performance degradation is in-line with the microbenchmarks for
\texttt{openat()}, \texttt{read()}, and \texttt{write()} system calls.

\paragraph{Summary:} Network-based applications like lighttpd and Nginx perform
close to native results whereas file-based applications observe overheads
between 4 and 55\% depending on the test scenario.
Most impacted are applications which access small files like SQLite.
In comparison to ptrace-based techniques, \system\ outperforms by 38-
529\%.

%% file: use_eval.tex
%%----------------------------------
\subsubsection{Least-Privilege Nginx Performance}
\label{sec:boxeval}
%%----------------------------------

  \begin{figure}[t]
      \begin{tikzpicture}
      \pgfplotsset{
          width=6.8cm,
          height=4.3cm,
          every tick label/.append style={font=\footnotesize},
      }
      \begin{axis}[
      xlabel={file size [KB]},
      xlabel style={at={(0.5,-0.1)}, font=\footnotesize},
      ylabel={Normalized throughput},
      ylabel style={at={(-0.09,0.5)}, font=\footnotesize},
      xmode=log,
      log basis x={2},
      xmin=1, xmax=512,
      log ticks with fixed point,
      ymin=0.5, ymax=1,
      x tick label style={/pgf/number format/1000 sep=\,},
      y tick label style={font=\footnotesize},
      legend pos=south east,
      legend style={at={(1.25,0.9),font=\scriptsize},
          anchor=north,legend columns=1},
      ymajorgrids=true,
      grid style=dashed,
      ]
      \addplot[color=blue!70!white, mark=square]
      coordinates { (1,0.965)(2,0.965)(4,0.965)(8,0.965)(16,0.960)(32,0.962)(64,0.960)(128,0.953)(256,0.949)(512,0.935)(1024,0.928)};
      \addlegendentry{\srand\_16}
      \addplot[color=red, mark=star]
      coordinates { (1,0.962)(2,0.962)(4,0.962)(8,0.962)(16,0.958)(32,0.960)(64,0.958)(128,0.950)(256,0.945)(512,0.930)(1024,0.919)};
      \addlegendentry{\semph}
      \addplot[color=brown!70!black, mark=diamond]
      coordinates { (1,0.983)(2,0.983)(4,0.983)(8,0.983)(16,0.979)(32,0.980)(64,0.979)(128,0.970)(256,0.964)(512,0.948)(1024,0.936)};
      \addlegendentry{\sdemph}

      \addplot[color=magenta, mark=triangle]
      coordinates { (1,0.941)(2,0.940)(4,0.940)(8,0.940)(16,0.940)(32,0.936)(64,0.934)(128,0.925)(256,0.916)(512,0.896)(1024,0.876)};
      \addlegendentry{\sscet}
      \addplot[color=lime, mark=+]
      coordinates { (1,0.960)(2,0.960)(4,0.959)(8,0.959)(16,0.955)(32,0.955)(64,0.952)(128,0.943)(256,0.933)(512,0.912)(1024,0.892)};
      \addlegendentry{\sdcet}
      \addplot[color=darkgreen!60!white, mark=o]
      coordinates { (1,0.704)(2,0.704)(4,0.704)(8,0.704)(16,0.694)(32,0.689)(64,0.672)(128,0.637)(256,0.589)(512,0.522)(1024,0.450)};
      \addlegendentry{strace}
      \end{axis}
      \end{tikzpicture}
      \caption{\label{fig:safe_sand} Normalized throughput of privilege
      separated NGINX using TLS v1.2 with \texttt{ECDHE-RSA-AES128-GCM-SHA256,
      2048, 128}. Std. dev. below
    1.5\%.}
      % All the data is measured on 01/17/2021
  \end{figure}
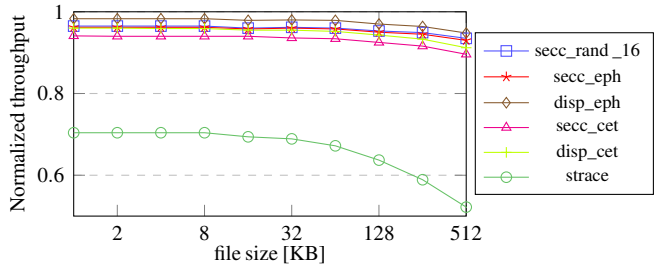

We evaluate the least-privilege Nginx shielding OpenSSL and sandboxing the HTTP
parser (see \S~\ref{sec:lp_nginx}). We measure the throughput downloading
files with varying sizes and normalize to the native performance (see
Figure~\ref{fig:safe_sand}).
Strace suffers from its interception costs and falls below 50\% for large files.
%----------------------------------
The results indicate less than 10\% overhead for the different \nexpoline
techniques and the lowest for \sdemph at 3-5\%.
The number of system calls scales with file size causing decreasing the
performance of \system.
Each \xcall consumes 269 cycles for CET-based \nexpoline (see
Appendix~\ref{app:cet_eval}) and 116 CPU cycles for random and ephemeral.
In addition to {\nexpoline}s, there are 89 {\xcall}s during the initialization
of OpenSSL, 16 to initialize each new HTTPS session and an increasing number of
{\xcall}s with increasing file size due to higher network bandwidth. (see
Table~\ref{tab:gate_count})

\begin{table}[t]
	\small
	\caption{\label{tab:gate_count} \xcall\ count for different file sizes in the
		test scenarios including startup of the process.}
	\begin{center}
		\begin{tabular}{ |c||c|c|c|c|c|c| }
			\hline
			File size & 1k & 4k & 16k & 64k & 256k & 1024k \\
			\hline
			Count & 129 & 129 & 132 & 141 & 177 &312 \\
			\hline
		\end{tabular}
	\end{center}
\end{table}

%% file: rw.tex
%%----------------------------------------------------------
\section{Related Work}
%%----------------------------------------------------------
%We survey background and related work about isolating memory within the
%application process, and virtualizing system calls and signals.
%
%-------------------------
%\subsection{Intra-process Isolation}
\label{sec:rw-intra-process}
%-------------------------
Traditional address space isolation provided by the operating systems is hardly
used for least privilege use cases.
Its limitation as described in \S~\ref{sec:motivation} outweigh the
benefits.
In recent years, browsers adopted process-based isolation to limit the blast
radius of the bug-prone renderer.~\cite{reis2019site}
Intra-process isolation research aims to decompose applications into domains,
compartmentalize each domain, and ensure domains cannot escalate their
privileges gaining access to another domain.
Their approaches can be classified into language-based, OS-based and
hardware-based techniques.
%--------------------------------
\paragraph{Language-based techniques}
Software Fault Isolation (SFI)~\cite{Wahbe:Efficient:1993} started a research
field to inline security checks within application code.
The goal is to translate an application via a compilation pass to enforce
security properties such as Control-Flow Integrity
(CFI)~\cite{abadi2005control-flow}, Code-Pointer
Integrity~\cite{Kuznetsov:Code:2014}, or intra-process
isolation~\cite{Koning:No:2017}.
Early techniques such as CCured~\cite{CCured:2005} and
Cyclone~\cite{jim2002cyclone} insert bounds checks at compile time to trigger
exceptions during the execution.
These techniques have, e.g., been used to protect Java applications from Java
Native Interface~\cite{tan2006safe, sun2012jvm, siefers2010robusta,
chisnall2017cheri}.
Similarly, web applications need protections from the native
code~\cite{Yee:Native:2009} which has lead to the development of Webassembly.
Webassembly has become the defacto standard for language-based intra-process
isolation.
These techniques suffer from non-negligible performance overheads and similarly
to \system have to also protect against attack vectors via the OS.
\system provides the \arch to generalize this abstraction and could also be
combined with language-based techniques, but we chose \intel MPK for its
superior performance.

%-------------------------------------
\paragraph{OS-based techniques}
To improve upon the limitations of existing process-based isolation, several
kernel abstractions have been suggested.
light-weight contexts(lwC)~\cite{Litton:Lightweight:2016}, secure memory
views(SMV)~\cite{Hsu:Enforcing:2016} and nested
kernel~\cite{Dautenhahn:Nested:2015} introduce a light-weight form of processes
allowing to associate multiple virtual address spaces with the same process
allowing for faster switches.
Shreds~\cite{Chen:Shreds:2016} combines OS-based and language-based techniques
to further improve the performance, and Wedge~\cite{Bittau:Wedge:2008a} combines
with runtime checks.
While these techniques improve the state of the art, their abilities are limited
by the provided hardware performance for privilege level switching.
To the contrary, \system takes advantage of intra-process isolation which avoid
these switches.

%-------------------------------------
\paragraph{Hardware-based}
In recent years CPU vendors have suggested techniques to bypass the costly
privilege level switch by, e.g., providing a VMFUNC
instruction~\cite{:Intel:2020} to switch address spaces in the CPU from
userspace or alter memory access permissions via userspace instructions like
WRPKRU.
Dune~\cite{Belay:Dune} uses \intel VT-x to elevate a process to run both in
privileged and unprivileged CPU mode.
This allows to use the system call boundary to efficiently decompose
applications and isolate components.
Secage~\cite{liu2015thwarting} achieves a similar separation using {\intel}'s
VMFUNC instruction which allows userspace applications to switch into a
different address space.
Koning \etal~\cite{Koning:No:2017} build a tool to compile applications with
varying isolation techniques and demonstrate the performance differences.
Generally, the overhead is dominated by the efficiency of the memory permission
switch technology or the language-based runtime security checks.
ERIM~\cite{Vahldiek-Oberwagner:ERIM:2019}, HODOR~\cite{Hedayati:Hodor:2019} and
Donky~\cite{Schrammel:Donky:2020} enable the secure use of \intel MPK for memory
isolation and demonstrate an efficient isolation technique.
Unfortunately, their techniques have several short comings as demonstrated by
Conner \etal~\cite{Connor:PKU:2020} which \system addresses without reverting
to inefficient isolation techniques keeping the premise of MPK-based isolation
techniques.
The general applicability of MPK-based techniques is demonstrated by
FlexOS~\cite{lefeuvre2021flexos} and Sung et al.~\cite{sung2020intra} which use
MPK in a unikernel environment.

\paragraph{System Call and Signal Virtualization}
%-------------------------
Pitfall~\cite{Connor:PKU:2020} demonstrated how the operating system can be used
to bypass intra-process isolation as discussed in the previous section.
To prevent such bypass, all operating system interactions need to be mediated,
strict security policies enforced, and then virtualized.
This style of attack elevated the importance of system call monitoring and
virtualization while also demonstrating that existing kernel-based techniques
are insufficient.

%-------------
Linux security module(LSM)~\cite{Wright:Linux:2003} intercept system calls in
the kernel and can be used to implement a system call filter as suggested by
SELinux~\cite{Loscocco:SELinux}, AppArmor~\cite{Mick:AppArmor},
Tomoyo~\cite{Harada:Tomoyo} and Smack~\cite{Schaufler:Smack}.
However, LSM would have to be extended to recognize different domains in
userspace and its interface does not support modifying system call arguments
which makes virtualization infeasible.
%
\iffalse
Similar to LSM are techniques such as Janus~\cite{Goldberg:Janus} which utilizes
\Code{procfs} to monitor and intercept the system calls in user space and filter
out not allowed system calls.
\fi
Alternatively, Seccomp~\cite{Corbet:Seccomp:2009} offers the userspace a
programmable system call filter using eBPF programs.
Seccomp is widely used by numerous applications such as Native
Client~\cite{Yee:Native:2009} to limit the OS interface accessible to the
untrusted component to a bare minimum.
Seccomp cannot easily be extended to filter different domains.
\system in some configurations relies on Seccomp and works around its
limitations.
%
\iffalse
Other applications like Ghavamnia et al~\cite{Ghavamnia:Temporal}, and
sysfilter~\cite{Demarinis:sysfilter} use static and dynamic analysis to provide
filtering policy on compile time or runtime.
%
But unfortunately, the system call filtering techniques only support filtering
of the system call execution, it does not emulate or virtualize the system call
itself.
\fi
%--------------------------
\if 0
System call을 intercept한 후 이를 가상화하는 어프로치로 대표적인 다음의 연구들이 있다. 우선 ptrace를 활용하는 것이다. ptrace는 다른 process에 attach해서 메모리 및 system call을 모니터할 수 있는 system call이며 이를 활용하여 syscall virtualization에 활용할 수도 있는데, ~~와 같이 tracing을 통하여 malware를 검출하고자 하는 방법들이 있고, 태수처럼 intercept하여 policy를 enforcing하는 기법들이 있으며 ERIM에서도 ptrace를 사용하여 mmap과 같은 memory sysscall을 가상화하였다. 또한 ptrace를 사용하지 않고 커널을 수정하여 유사한 기능을 제공하는 ostia도 있다. 그러나 ptrace 방법들은 여러 프로세스 사이에서 커널을 거쳐서 동작하기 때문에 퍼포먼스가 매우 낮아진다는 단점이 존재한다.
\fi

To virtualize the system calls, there are a number of efforts utilizing
\Code{ptrace()} system call.
\Code{ptrace()} allows a process to attach another process and monitor the
memory and system calls and it is widely used for the debugging and the
profiling purpose like strace utility.
\iffalse
%
There are efforts to detect malware based on the system call tracing by ptrace
such as Jain et al.~\cite{jain2000user}, and
DroidTrace~\cite{zheng2014droidtrace} provides dynamic analysis toolfor Android
by utilizing ptrace.
%
\fi
ERIM~\cite{Vahldiek-Oberwagner:ERIM:2019} use ptrace to intercept memory-related
system calls such as \Code{mmap()} and virtualize the system calls to fulfill
with the memory policy.
%
\iffalse
Other techniques like Ostia~\cite{garfinkel2004ostia} modified the kernel to
monitor and delegate the system calls to a monitor process and the monitor
process emulate the system calls.
\fi
%
However, \Code{ptrace()} approach is known to suffer from the high performance
overhead because the system call interposition requires multiple context
switches.

%% file: conc.tex
\section{Conclusion}
%%----------------------------------------------------------

  % --------------------
  %Nested privilege separation has made great strides but neglected several
  %design and implementation level issues.
  %
  This paper introduces \arch, which builds a new virtual machine abstraction
  for representing subprocess authority.
  A self-isolating monitor efficiently enforces the abstraction by mapping the
  subprocess authority to system level objects.
  We demonstrate the usefulness of \arch by decomposing Nginx to enforce
  least-privilege for its HTTP parser and OpenSSL library.
  Our prototype, \system, explores implementation tradeoffs and shows that
  sub-process authority can be enforced efficiently (within 5\% of native
  performance).
  \iffalse
  %
  System virtualization allows each \domain to use the full set of kernel
  features while remaining fast and secure.
  %
  \system achieves this goal by 1) controlling all system access through a
  novel, lightweight runtime monitor, 2) systematically enforcing information
  flow policies, and 3) exporting a new privilege separation abstraction for
  easy privilege separation.
  %
  It improves performance by working solely at userlevel and simplifies system
  object flow security by enforcing at the \syscall level.
  %
  Our results indicate that it is possible to meaningfully enhance security of
  nested privilege separation, while enabling new separation abstractions.
  \fi

%% file: mpksep-bg.tex
  %%--------------------------------------------------------
  \section{Intra-Process Memory Isolation}
  \label{sec:appmem}
  %%--------------------------------------------------------
  % --------------------
  %\subsection{Nested Privilege Separation}
  
  This appendix describes the background of intra-process memory isolation, including the abstractions and pitfalls of memory protection keys, and the systematic abstractions needed for a comprehensive memory isolation mechanism.

  %%--------------------------------------------------------
  \subsection{Protection Background}
  %%--------------------------------------------------------
    % --------------------
    Hsu \etal~\cite{Hsu:Enforcing:2016} describe three generations of privilege
    separation, each increasing from manual, address-space isolation to the
    third generation that efficiently enables concurrent per-thread memory
    views.
    % of \segments using an object model~\cite{Lampson:Protection:1974}.
    %
    The key is new hardware that extends paging with userlevel tags for fast but
    insecure isolation.
    %
    \iffalse
    %
    ensure isolation at a single privilege level and intra-address space.
    %
    The biggest challenge and cause for progress has been performance costs of
    separation mechanisms.
    %
    To deal with this recent hardware support has explored mechanisms that trade
    of security for performance by placing the isolation mechanism in the
    application itself.
    %
    This means that \domain switches can occur quickly, but comes at the cost of
    exposing the protection data to compromise.
    %
    %The unique challenge of doing so within a single address space is that the
    %protection mechanism is nested within the space itself, thus making it
    %accessible to the runtime. 
    %
    \fi

    %%--------------------------------------------------------
    \paragraph{\intel Memory Protection Keys (MPK)}
    %%--------------------------------------------------------
    % Point: How does MPK work?
    %
    MPK~\cite{:Intel:2020} extends page tables with a 4-bit tag for labeling
    each mapping.
    A new 32-bit CPU register, called \pkru, specifies access control policies
    for each tag, 2-bits per for controlling read or write access to one of the
    16 tag values.
    The policy is updated via a new ring-3 instruction called \wrpkru.
    On each access, the CPU checks the access control policy as specified by the
    mapping's tag and associated policy from the \pkru.
    If not permitted the CPU faults and delivers an exception.
    %
    %This access control happens in addition to existing PT policies, operating
    %as an orthogonal or hybrid protection policy.
    %
    %As a result, these access permission checks do not decrease the performance
    %of an application.

    %%--------------------------------------------------------
    \paragraph{MPK Security vs Performance}
    %%--------------------------------------------------------
    % Point: indicate how MPK trades off security for
    % performance
    % --------------------
    Unfortunately, the \pkru can be modified by any userlevel \wrpkru
    instruction: MPK is bypassable using gadget based attacks.
    %
    %so as delivered by real hardware (\eg Intel MPK), it is not possible to
    %defend against all attackers, nor was that the original threat model.
    %
    As such MPK balances security and performance by allowing protection changes
    without switching into the kernel.
    %was to develop a protection mechanism that could be used isolation without
    %requiring a call into the kernel, which would be required to use page table
    %based separation.
    %
    %Unfortunately, this means that even if the system is privilege separated,
    %an attacker can disable security by modifying the \pkru through hijacking
    %the \wrpkru instruction.
    %
    %But, is it possible to restrict usage of the \wrpkru instruction to ensure
    %guaranteed isolation?  Making it guaranteed for a nested secure monitor
    %requires an additional piece of machinery.

    %%--------------------------------------------------------
    \paragraph{Preventing MPK Policy Corruption}
    %%--------------------------------------------------------
    % Point: Overview method for preventing PKRU corruption,
    % don't focus too much on details push to later
    % --------------------
    Nested privilege separation reconciles the exposure of protection state by
    ensuring \wrpkru instructions are only used safely by the \monitor.
    They achieve this by removing all \wrpkru instructions from the untrusted
    binary and crafting \emph{nested call gates} that prevent
    abuse~\cite{Vahldiek-Oberwagner:ERIM:2019, Hedayati:Hodor:2019, Enclosure,
    Chen:Shreds:2016, sung2020intra, lefeuvre2021flexos}.
    %
    \iffalse % not sure need the ref here
    %
    While the idea has been applied to operating
    systems~\cite{Wahbe:Efficient:1993, Dautenhahn:Nested:2015,
    Song:Enforcing:2016, Criswell:Secure:2007, Criswell:Virtual:2014,
  Criswell:KCoFI:2014, Azab:SKEE:2016},
  hypervisors~\cite{Shi:Deconstructing:2017}, and trusted execution
  environments~\cite{Hua:vTZ:2017}, ERIM~\cite{Vahldiek-Oberwagner:ERIM:2019}
  and HODOR~\cite{Hedayati:Hodor:2019},
    %and Donky~\cite{Schrammel:Donky:2020}
    have done this for user mode separation.
    %
    \fi
    %
    % NDD: Do we need to mention donky here?
    %Donky suggests hardware modifications to prevent exploitation of MPK-like
    %mechanisms.
    %
    %While in hardware, Donky still requires a system virtualization framework
    %that could be efficiently provided by \system.
    %
    %maintain their fast isolation while securing them by virtualizing system
    %interfaces.

    % --------------------
    %Recent work employs a software technique that
    %\emph{virtualizes} privilege within a process by using a
    %combination of commodity hardware enforcement mechanisms
    %along with a technique we call \emph{instructions as
    %capabilities}.
    %
  
    \input{memiso}

%% file: memiso.tex
  % --------------------
  \subsection{Privilege and Memory Virtualization}
  \label{sec:p_erim}
  % --------------------

    % --------------------
    In this section, we provide an overview of
    ERIM~\cite{Vahldiek-Oberwagner:ERIM:2019},
    which we extend to build \system.
    We encourage the reader to review detailed methodology
    from the original work.
    
    %%--------------------------------------------------------
    \paragraph{Thread Model}
    %%---
    An initial configuration partitions the application
    into the \td and \ud, where the \td contains the trusted
    monitor and the \ud contains the rest.
    Once the application is separated so that the parts are differentiated, the
    system is configured so that all pages of the \td have key \Code{0} and \Code{1}
    based on the confidential requirement, and all pages of the \ud have key \Code{2}.
    Some pages will have other keys if they belong to other subdomains in \ud.

    %%--------------------------------------------------------
    \paragraph{Virtual Privilege Switch}
    %%--------------------------------------------------------
    One of the most important elements when nesting the \monitor into the same
    address space is the need for secure context switching, which is complex to
    get correct because an attacker has access to whatever is mapped into the
    address space.
    %
    %For example, one of the basic properties we ensure is to only allow \td to
    %directly execute \syscall operations, but if they are both mapped in and
    %the \td is left executable for performance reasons, then the \ud could
    %\Code{jmp} to it.
    %
    While the \td is executing the \pkru is configured to \Code{allow\_all}
    (read/write to all domains), and operating in the \td virtual privileged
    mode.
    While the \ud is executing the \pkru value for the key \Code{0} will be
    \Code{deny\_write} and \Code{deny\_all} for \Code{1}.
    The \emph{virtual domain switch} is implemented as a change in the
    protection policies in the \pkru---when entering the monitor set the policy
    to \Code{allow\_all}, when exiting restore the original key based on the
    previous state.
    This means that whenever the value of \pkru changes so too does the
    currently executing domain.
    Each entry point into \system is setup with a call gate with a \wrpkru that
    transitions the domain.
    The basic idea is to nest monitor code directly into the address space of
    the application and wrap each entry and exit point with a \wrpkru operation.
    By doing this the system can transition between contexts and only allow
    monitor code to access protected state---a virtual privilege switch.
    This similar technique is also used to switch between different subdomains
    to enable the usage of other keys in \ud. 

    %%--------------------------------------------------------
    \paragraph{Securing the Domain Switch}
    % --------------------
    Unlike systems with real hardware gates, this SW/HW
    virtual privilege switch has challenges because the
    instruction must be mapped as executable to allow fast
    privilege switching.
    The first thing an attacker could do is use a direct
    jump to any code in the monitor and thus bypass the
    entry gate.
    This would in fact allow the attacker to execute monitor
    code.
    One way to thwart could be to modify the executable
    policy on the monitor pages, but that would require a
    call into the OS which defeats the purpose of fast
    domain switching of MPK in the first place.
    Instead, we observe that even if an attacker is able to
    jump into the middle of the monitor the domain would
    have never switched, therefore, none of the protected
    state is available for access and therefore the basic
    memory protection property holds.
    The only way to change the domain is to enter through
    the entry gate.
    %

    %%
    %I think we should discuss this here.. instead of \ref{gentry},
    %but ... it should be that complete.. so ..

    Since the switch is a single instruction, we can easily 
    verify the result of such switching immediately after 
    the {\tt WRPKRU} instruction and loop back if it is not 
    switch to the intended PKRU state. This ensure that the 
    PKRU state at all exits of the gate secquence will be the 
    intented PKRU state. 
    %%%

    %
    Effectively, the attacker now faces the dilemma that jumping
    into the middle of the code will ended nothing since it is 
    the equivelent of running the same code in any other locations, 
    or it can try to jump to the entry gate, but any landing places of 
    the gate will only switch to the correct PKRU value and continue 
    the exeuction with deterministic control flow. No code can be abused.

    %%--------------------------------------------------------
    \paragraph{Instruction Capabilities}
    % --------------------
    % --------------------
    Alternatively, an attacker could generate their
    own unprotected variant of \wrpkru---if an attacker can
    inject or abuse the \wrpkru instruction, they could
    switch domains and gain access to the monitors protected state.
    To deal with this ERIM and others like it use a
    technique called \emph{instruction capabilities}: that
    is by using a combination of static transformations and
    code validation and dynamic protections an instruction
    becomes much like a capability.
    The static analysis removes all instances of the \wrpkru
    opcode so that the attacker has no aligned or unaligned
    instructions that could write the value without
    monitoring, and dynamic runtime is configured so that
    all code is writeable or executable but not both.

    % --------------------
    \paragraph{Controlling mode switches}
    % --------------------
    % --------------------
    Processes may switch into 32-bit compatibility mode,
    which changes how some instructions are decoded and
    executed.
    The security monitor code may not enforce the intended
    checks when executed in compatibility mode.
    Thus, we insert a short instruction sequence immediately
    after WRPKRU or XRSTOR instructions that will fault if
    the process is in compatibility mode.
    See Appendix~\ref{app:32bit-mode} for more details.

    \iffalse
  % --------------------
  \paragraph{Guaranteed Protected Entry}
  % --------------------
\label{gentry}
    % --------------------
    One of the basic properties is that the \ud cannot jump
    into the middle of the protected subsystem.
    %
    A naive solution to this would be to map all \td code as
    non-executable whenever control flows to the \ud,
    however, this would require a call into the kernel to
    modify the Page Tables and thus eliminate much of the
    performance boost of intra-process protection.
    %
    Instead, we develop a technique that allows indirect
    jumps into the \ud but makes each case harmless.
    %
    The main cases of problem are the following:
    %
    1) jump anywhere,
    %
    2) abuse privilege instructions to disable protection
    and gain control,
    %
    3)

    \fi

%% file: sec_eval_apdx.tex
%%----------------------------------------------------------
\section{New Attack Vectors and Security Evaluation} \label{sec:apdx_sec_eval}
%%----------------------------------------------------------

In addition to the attacks described by Conner \etal~\cite{Connor:PKU:2020}, we
found several attacks against intra-process system call and signal
virtualization.
For the evaluation, we created a fixed address secret inside trusted domain.
All test cases try to steal this secret and hence, would break \system's
isolation guarantees.
The attacks try to bypass \system by performing system calls modifying the
protection policy of the secret or trying to elevate itself to be trusted by
overriding the PKRU register.
They specifically target the implementation of \system and highlight the degree
to which \system has followed through with its security guarantees.
Table~\ref{tab:qsa} summarizes the the results.

\paragraph{Forged Signal} \system effectively prevents the basic {\tt sigreturn}
attack from~\cite{Connor:PKU:2020}.
However, the kernel places signals on the untrusted stack and delivers the
signal to our monitor signal entrypoint.
The untrusted application may forge a signal frame and directly call the
monitor's signal entrypoint.
As a result, it can, e.g., choose the PKRU value and the return address.
Therefore, the entrypoint has to distinguish between a fake signal from the
untrusted application or a real signal from the kernel.
The entrypoint is carefully constructed such that a signal returning from kernel
returns with privileges from the trusted domain and hence, is capable of writing
trusted memory.
We rely on this observation and place an instruction at the beginning of the
monitor which raises a flag in the trusted monitor.
Any fake signal created by the untrusted application cannot raise the signal
flag in trusted memory which violates a check that cannot be bypassed in the
monitor's signal entrypoint.

\paragraph{Fork Bomb} This attack targets the random location of the system call
instruction in \system.
To perform a system call the untrusted application may guess the random location
of the system call instruction.
Assuming the trampoline size is 16 pages, there are 65534 possible locations of
the system call instructions.
When the untrusted application is capable to fork children, the untrusted
application may try different locations within each child.
In case the child crashes, the system call was unsuccessful and the untrusted
application has to retry.
Using this brute force algorithm the untrusted application tries until a child
does not crash.
At this point the untrusted application has access to a child process that
bypassed {\system}'s security guarantees and may perform arbitrary system calls.
It should be noted that only \srand is successive to this
attack, since \semph removes the system call instruction completely
when returning control to the untrusted application.

\paragraph{Syscall Arguments Abuse} \system virtualizes a subset of all system
calls.
System calls which are not virtualized could be exploited to read secret memory,
unless \system verifies that all pointers provided to a system call lie within
untrusted memory.
We perform an attack based on the {\tt rename} system call and pass it a memory
pointer from the trusted domain as an argument.
\system successfully prevents this attack by checking the pointer locations.

\paragraph{Race Condition in Shared Memory Access} Shared memory may be used across
multiple processes to bypass {\system} checks on arguments to system calls.
In particular, we consider a {\tt pwritev}-based attack in which a child process
performs a {\tt pwritev} system call using an IO vector in shared memory.
If the parent was permitted access to the same shared memory, it could time to
alter the IO vector's values to point to trusted memory.
This attack has to be timed such that the child's monitor has already performed
the security checks, but the system call has not yet read the affected IO memory
vector.
\system prevents such attacks by copying pointers in system call arguments to
the trusted memory region and only then performing the system call using the
copied arguments.

%------------------------
\paragraph{Race Condition in Multi-Threaded Applications}
Supporting multi-threading is essential in modern computing environment that
\system also supports it.
But, there are a few attack surfaces which use race conditions in multi
threading environment.
First, indirect jump to \sysretg is possible in ephemeral \system.
For example, one thread calls a \syscall\ which take very long time, and the
attacker thread jumps to the active \sysretg.
To prevent such attacks, we use either Syscall User Dispatch, or per-thread
Seccomp filter.
Second, the attackers could perform TOCTOU attacks in the \syscall\
virtualization.
For example, one thread open a normal file and call a file-backed \syscall,
while another thread close the file descriptor and open a sensitive file which
is not allowed for the untrusted code.
In \system, we provide locks per file descriptor that \texttt{close} system
call could be locked when another thread is using that file descriptor.
As well, \system provides a lock for memory management \syscalls\ and signal
related \syscalls.

%Beside these tests, we also have some test cases to verify the behavior of our
%implementation. For example, creating a new thread should fail or using
%{\tt mprotect} to create an executable shared memory should fail.

\paragraph{TSX Attack} TSX is an extension to support transactional memory in x86.
It has a similar principle as exception handling but at the hardware level. When
any considered as a violation of transactional happens, the hardware rollback and
modification and jump to a preset restore code.
Unfortunately, because the rollback feature provides a harmless way of content probing
since the first introduction, it has been used as a source of memory leakage.
It has been obsoleted in the latest \intel CPU but still exists in many products with MPK.
Our attack utilizes TSX as a probe to the randomized trampoline. First, a \Code{xbegin}
is used to enable the TSX environment. Then, we call an address within the trampoline
region. Now, there are three cases about the content on target address, \Code{int3},
\Code{syscall} and \Code{ret}.
For the first two cases, TSX will be aborted but in the second case, \Code{ret} instruction
can be executed successfully. Such difference is sensible from the view of the attacker and
the address contains \Code{ret} is exposed. Because our syscall gadget is \Code{syscall;ret}.
This exposed the secret address of \Code{syscall}.
Fortunately, TSX can be disabled through kernel or BIOS, and among all \system configurations,
only \srand is secret-based and susceptible.

%% file: ceteval.tex
%%----------------------------------------------------------
\section{Analyzing CET}\label{app:cet_eval}
%%----------------------------------------------------------

  % --------------------
  The experimental results for CET indicated much higher overheads than we
  anticipated, and so we explored CET under different scenarios to understand if
  the overheads were truly CET or if there might be something in the Linux
  implementation or \system.
  To understand the details we perform a minimal experiment on LMBench, Lighttpd,
  and Nginx.

 We report on three configurations: 1) no CET at all as baseline, 2) CET is
 enabled in the benchmark application binary and the dependent shared libraries,
 and 3) CET are enabled all the dependent binaries including glibc.

 The result of the tests says that both CET enabled configurations have
 2-8\% of overhead compare to the baseline configuration, that is, only enabling
 CET induces at most 8\% overhead.
 In addition, IntraVirt performs multiple domain management, which contains
 multiple shadow stack management and shadow stack switch on domain switch, therefore
 CET enabled version of IntraVirt has more overhead than others.

 Examining CET performance is a key element of related work but was out of
 scope for the analysis at this time.

%% file: app-32bit-mode.tex
\section{Controlling mode switches}
\label{app:32bit-mode}

This appendix explains why it is necessary to check that the application is executing in 64-bit mode when it enters
the trusted code, and it also describes a mechanism for performing that check.

64-bit processes on Linux are able to switch to compatibility mode, e.g. by performing a far jump to a 32-bit code
segment that is included in the Global Descriptor Table (GDT).  Executing code in compatibility mode can change
the semantics of that code compared to running it in 64-bit mode.  For example, the REX prefixes that are used to
select a 64-bit register operand size and to index the expanded register file in 64-bit mode are interpreted as
\texttt{INC} and \texttt{DEC} instructions in compatibility mode.  Another example is that the RIP-relative addressing mode in
64-bit mode is interpreted as specifying an absolute displacement in compatibility mode.

Executing the trusted code in compatibility mode may undermine its intended operation in a way that leads to
security vulnerabilities.  For example, if the trusted code attempts to load internal state using a RIP-relative
data access, that will be executed in compatibility mode as an access to an absolute displacement.  The
untrusted code may have control over the contents of memory at that displacement, depending on the memory
layout of the program.  This may lead to the trusted code making access control decisions based on forged
data.  Conversely, if the trusted code stores sensitive data using a RIP-relative data access, executing the
store in compatibility mode may cause the data to be stored to a memory region that can be accessed by
the untrusted code.

To check that the program is executing in 64-bit mode when it enters the trusted code, a sequence of
instructions such as the following may be used:
\begin{enumerate}
\item Shift \texttt{RAX} left by 1 bit.  In compatibility mode, this is executed as a decrement of \texttt{EAX} followed by a
  1-bit left shift of \texttt{EAX}.
\item Increment \texttt{RAX}, which sets the least-significant bit of \texttt{RAX}.  In compatibility mode, this first decrements
  \texttt{EAX} and then increments \texttt{EAX}, resulting in no net change to the value of \texttt{EAX}.
\item Execute a \texttt{BT} (bit test) instruction referencing the least-significant bit of \texttt{EAX}, which is valid in both 64-bit mode
  as well as compatibility mode.  The \texttt{BT} instruction updates \texttt{CF}, the carry flag, to match the value of the
  specified bit.  It does not affect the value of \texttt{EAX}.
\item Execute a \texttt{JC} instruction that will jump past the next instruction iff \texttt{CF} is set.
\item Include a \texttt{UD2} instruction that will unconditionally generate an invalid opcode exception, which
  will provide an opportunity for the OS to terminate the application.  The security monitor should prevent
  the untrusted code from intercepting any signal generated due to invalid opcode exception from this
  code sequence.
\item Shift \texttt{RAX} right by 1 bit to restore its original value.  This instruction is unreachable in compatibility mode.
\end{enumerate}
The preceding description of the operation of the instructions in compatibility mode assumes that the default
operand size is set to 32 bits.  However, a program may use the \texttt{modify\_ldt} system call to install a code segment
with a default operand size of 16 bits.  That would cause the instructions that are described above as accessing
\texttt{EAX} to instead access \texttt{AX}.  That still results in the instruction sequence detecting that the program is not
executing in 64-bit mode and generating an invalid opcode exception.  Furthermore, \system can block
the use of \texttt{modify\_ldt} to install new segment descriptors.  None of the default segment descriptors in
Linux specify a 16-bit default operand size.

It is convenient to use \texttt{EAX/RAX} in the preceding instructions, because the REX prefix for accessing
\texttt{RAX} in the instructions used in the test happens to be interpreted as \texttt{DEC EAX},
which enables our test to distinguish between
64-bit mode and compatibility mode as described above by modifying the value of the register that is
subsequently tested in the \texttt{BT} instruction.  However, we need to restore the value of \texttt{EAX/RAX}
after the mode test.  One option would be to store \texttt{RAX} to the stack.  However, that may introduce a
TOCTTOU vulnerability if the untrusted code can modify the saved value.  That is why we used shift operations
to save and restore the original value of \texttt{RAX}, depending on the property that only the least-significant
32 bits of RAX are ever set at the locations where mode checks are needed.

The mode test comprises 11 bytes of instructions total.

The mode test instruction sequence overwrites the value of the flags register.  If the value of the flags register
needs to be retained across the mode test, that can be accomplished using a matching pair of \texttt{PUSHF} and
\texttt{POPF} instructions surrounding the mode test.  These instructions are encoded identically in 64-bit mode and
compatibility mode.  It may be possible for untrusted code to overwrite the flags register value while it is saved
to the stack.  However, trusted code should not depend on flags register values set by untrusted code,
regardless of whether that register has been loaded from stack memory or it has been set by the processor
directly as a side-effect of executing instructions in untrusted code.

If the instruction sequence for testing the value of \texttt{EAX/RAX} used with an \texttt{XRSTOR} or
\texttt{WRPKRU} instruction that is not followed by trusted code is valid in all modes that are reachable
by the untrusted code, then the mode test code may be omitted prior to that value test code.

%% file: archive-flow.tex
\section{System Flow analysis}

\if 0
  % --------------------
  The goal is to map the segment isolation policies to the flows through system
  objects.
  %
  In this section we first do an analysis of threats and group system calls into
  classes for protection.
  %
  We are basically creating a semantic mapping of the segment isolation policy
  to the system call interface so that we can monitor system calls and apply
  enforcement right there---letting us have a cool \syscall based monitor with
  distributed properties and no kernel mods.
\fi

  % --------------------
%\ndd{dumped from otehr section}
  This appendix describes the effective security policies of \system to prevent information flows via system objects.
  %One of the most unique aims of our work is to close the gap of information
  %flows through system objects.
  %
  The set of low-level \domain specific rules combined with system object
  monitoring enables a simple and default information flow policy of deny all
  information flows from a protected object to an object outside of its domain.
  For example, if one domain requests mmap of a region that it does not have the
  capabilities for the mmap will be rejected.
  In the next section we detail an initial separation with two domains that puts
  this abstraction on display and analyzes all system calls to identify all that
  lead to information flows that bypass the basic \domain property: \ie by
  default a domain is restricted to access only those things it has capabilities
  for.

%While previous work neglected information flow between system objects, this
%section's goal is to develop effective security policies preventing information
%flow across \domains via system objects.
It is important to find all
possible attack surfaces by listing up the system resources provided by the
operating system. In this section we first analyze the threats and group system
calls into classes.
Recent work, PKU pitfall~\cite{Connor:PKU:2020}, describes a number of attacks
toward ERIM~\cite{Vahldiek-Oberwagner:ERIM:2019} and
HODOR~\cite{Hedayati:Hodor:2019}, which are intra-process isolation
abstractions, but this work fails to analyze the threat more systemically and
does not build a complete set of security policy.

At first, the most simple and the most effective threat is to access the memory
of other domains. The attacker could simply directly access any virtual address
spaces in the same process, or she could use a vulnerable interfaces provided by
the system or other domains. Therefore, all the interfaces which accesses the
memory should be carefully designed and implemented.

In Linux, the file descriptor is shared within the process boundary. The
attacker could easily scan the opened file descriptor of the victim process by
scanning \texttt{/proc/self/fd} and the attacker is also free to control the
file descriptor as well. Therefore, the attacker could simply access the content
of the file or the socket, close the file descriptor and open another one with
the same file descriptor number, or move the offset that the victim domain could
access incorrect position of the file.

In addition, the code segment of each domain has to be protected. If not, the
attacker is able to jump to the code segment at any time by simply\texttt{jmp}
or \texttt{call} instructions, or by using more sophisticated attack techniques
like Return Oriented Programming. Therefore the system has to provide the
protection of this type of attacks, such as SFI and CFI.

Linux provides various special interfaces for exception handling, debugging and
profiling. For example, signals are used for handling any special occasion of
the process, and ptrace for debugging and profiling from other processes. These
interfaces let the kernel shares the control of the process and provides various
convenient functionalities, but it also provides the convenience to the attacker
as well. The attacker could perform the memory access, code execution, and
control the control flow of the process freely with these interfaces. Therefore,
Linux provides various protection mechanisms of such interfaces such as
YAMA~\cite{:YAMA:}, but they are all bounded to the process.

Linux also provides special interfaces to access the memory of the process
such as \texttt{/proc/self/mem}, which maps the virtual memory address of the
process into a file. The attacker could simply use the well known, fully allowed
file interfaces to access such an important file and access the protected memory
at any time.

Lastly, while each system call has to be considered in isolation, it is
important to also investigate potential threats due to concurrency leading to
time of check, time of use attacks (TOCTOU). For example, an attacker may
attempt to read a file while at the same time seeking to a protected region in
the file. Depending on the interleaving policies enforced for the read system
call may consider the file descriptor position before the seek, but the read
ultimately returns data from the new seek location.

  % --------------------
  \subsection{System Flow via System Calls}
  % --------------------

  %\todo{think about merging this and the previous section to shorten and simplify analisys}

\if 0
    % --------------------
    We systematically evaluate all the system calls in Linux to identify a set
    of basic potential flows.
    %
    Our results are consistent with prior attacks~\cite{PKRUpitfalls} while
    expanding to \todo{write expansions}.
    %
    In our analysis we found X types of ways a piece of data can flow and
    categorize them as: \todo{copy from bumjin's lists?}
    %
    \todo{any thing else to show our analysis... probably a table?}
\fi

We systematically evaluate all the system calls in Linux to identify a set of
basic potential flow and the policy. Our results are consistent with prior
attacks~\cite{Connor:PKU:2020} while expanding to other factors that the system
calls could affect. We list up the type behaviors of the system calls and
categorize them as the policy enforcement requirement.

At first, the group that the most system calls are the system calls which access
resource handles such as file descriptor or shared memory. Theses handles,
provided by kernel, are free to share in the process boundary that as mentioned
in \S\ref{sec:sysflow}, within the process, such handles are not
protected between domains and the attack could make use of the handle to affect
other domains. As well, depending on the threat model, the original resource
could be important as well as the handle itself. For example, some policy could
enforce only the opener of the handle could access the handle, or in some other
policy, the handles which points to the special resources such as
\texttt{/proc/self/mem} could be treated as a special handle in the policy.

The next category is the system calls which control the memory status of the
process and access the memory, such as \texttt{mmap()}, \texttt{mprotect()}, and
\texttt{brk()}. These group of system calls controls the memory map of the
process, change the permission, and sometime modify the memory contents, so it
is critical for the security. For example, the attacker could write code into a
read/write memory page, then change the permission to executable, then jmp to
the code. However, these system calls are considered to be essential for all the
applications, it is not possible to simply disallow them to the users.
Therefore, these type of system calls are carefully identified, analyzed, and
enforced by the policy.

There is a group of system calls, which controls the process flow, such
as \texttt{rt\_sigaction()}, \texttt{prctl()}, and \texttt{ptrace()}.
These system calls are able to control the flow of the process, access
the memory, access to kernel, and provided various different and powerful
interface with the kernel.
These type of system calls have to be carefully analyzed, and the policy
has be very sophisticated, because we cannot simply not allow these type
of valuable, and well known system calls which are used in many applications.
We will discuss signals in this paper, one of the most important example in
this group.

Next, we have a group of system calls which controls the hardware, such as
\texttt{pkey\_mprotect()} and \texttt{arch\_prctl()}.
These type of system calls mainly accesses the registers or hardware specific
resource, we could consider them as special system calls and we could disallow
them if necessary.

The last group is the system calls which controls the privilege of the resources,
such as \texttt{setuid()}, and \texttt{chown()}.
These type of system calls are mainly controlled by Linux Discretionary access
control mechanism and mostly requires higher privilege to be executed, we
are not specifically consider this group as important category in this work.

But unfortunately, each system call does not categorized as just one group.
For example, \texttt{read()} system call could access the file descriptor,
read the file contents, and as a biproduct, it could change the file offset
position.
But it does access the memory as well by the read buffer input parameter.
The attacker could execute \texttt{read()} to read a legitimate file, but
contains malicious code, and use readable and executable page of other domain
address for the read buffer, which will make the kernel simply overwrite the
address. In the same sense, \texttt{mmap()} could allow file access without
file related system calls by putting file descriptor in the input parameter.

As a result, each system call cannot be simply analyzed by its own behavior.
We have to analyze all the possible interfaces, including behavior, input
parameters, output parameters, and return values to properly derive the policy.

%% file: app-sig.tex
\section{Signal Virtualization with Untrusted Stack}
\label{sec:appsig}
    % Nathan stupid simple list
    % 1) pkru == semi trusted UT --> require signal proxy through IV,
    % 2) sigstate expose pkru -> require sigstate virt,
    % 3) need to deliver on U stack --> requires switch to T (wrpkru) first to push to U stack,
    % 4)  opens spoof attack because of wrpkru and no way to tell --> requires something,
    % 5) what about reentry while in IV cause bugs -> requires concurrency control no reentry,
    % 6) but no reentry breaks compat -> requires deferring/sigpending,
    % 7) what if signal delivered while in IV, can't deliver right away because U need syscall to finish -> requires sign pending and masking to simplify

    % --------------------
    Signals modify the execution flow of a process by
    pushing a signal frame onto the process stack and
    transferring control to the point indicated by signal
    handler.
    The primary reasons we must fully virtualize signals are
    because 1) Linux always resets \pkru to a
    semi-privileged state where domain 0 is made
    RW-accessible and all other domains are read-only and 2)
    because signals expose processor state through
    \Code{struct sigframe}, potentially leaking sensitive
    state or allowing corruption of \pkru, which could lead
    to \ud control while in the \td context. % (invariants
    %I\ref{inv:sigintegrity} and I\ref{inv:memiso}).
    %
    As such, \system must interpose on all signal delivery to
    minimally transition protection back to the \ud mode and
    virtualize signal handler state to avoid leakage and
    corruption.

    % --------------------
    \system accomplishes this by virtualizing signals so that all signal
    handlers are registered with \system first, and second, registering signals
    with the kernel so that \system always gains control of initial signal
    delivery.
    When a signal occurs \system first copies the signal
    handler context info to protected memory so that the \ud
    cannot read or corrupt it.
    Next \system must deliver the signal to the \ud, but to
    do so it must 1) push the signal info onto the \ud stack and
    2) switch the protection domain to the \ud.
    Unfortunately, the semi-protected \pkru state does not
    map the \ud stack as writable, so \system first modifies
    \pkru so that it is fully in the \td and then pushes the
    signal information onto the \ud stack.
    Then \system transitions to the \ud mode, giving control to
    the handler registered in the first step.

    % --------------------
    The next challenge is that the domain switch into the
    \td places a \wrpkru in the control path, which can be
    abused by the \ud to launch a \emph{signal spoofing}
    attack.
    By spoofing a signal, the \ud could hijack the return
    path to its own code while setting PKRU to the \td.
    As such, \system must first add a mechanism to detect
    whether the signal is legitimately from the kernel or if
    it is from the \ud.
    Figure~\ref{fig:entrypoint} shows our approach that uses a special flag that
    resides in the \td as a proof of PKRU status before \wrpkru.
    This flag is allocated with key \Code{0}, so it is writable only if the
    signal handler is invoked by the kernel which reset PKRU to default.
    A spoofed signal handler invocation from the \ud would result in a
    segmentation fault that can be detected by the signal handler.
    %

    % --------------------
    \input{figs/sigentry.tex}

    % --------------------
    The next major issue is dealing with signals being
    delivered while the \system system call virtualization
    is working in the \td.
    This can cause bugs due to reentry,
    leading to potential security violations due to
    corrupted state.
    We must guarantee that our signal handler can only be
    invoked by the kernel once until we decide to either
    deliver or defer the signal and return to the corresponding
    state.
    % [[during syscall is the only case]]
    %
    The second problem arises out of the complex nature of
    adding \system in between the \ud and kernel, in the
    case where the signal is delivered during \system's
    handling of \syscall.
    Unfortunately, we cannot simply ignore either these
    signals because that would break functionality.
    %\system simply finishes cleanup and then delivers the
    %signal, however, In the case where the \td is
    %interrupted and the signal is asynchronously delivered
    %before the \syscall has finished creates a problem.
    %
    %The problem is that under normal circumstances a signal
    %is delivered at the completion of the system call,
    %however, in this case the \td has started but not
    %finish the \ud's system call, creating a
    %synchronization issue.
    %
    In this case, \system must defer the signal till after
    the \syscall is completed.

\input{figs/sigstate}
    % --------------------
    The solution to interrupted signal delivery is to
    emulate almost exactly like the kernel.
    As depicted in Figure~\ref{fig:state}, signals occurring
    while in the \td will be deferred by adding them to an
    internal pending signals queue and masking that
    particular type of signal in the kernel.
    The latter step is not necessary but pushes the
    complexity of managing multiple signals of the same type
    to the kernel.
    Once the current operation is completed, \system selects
    the last available signal that has not been masked by the
    user and delivers it.

    % --------------------
    Signals represent the most complex aspect of \system.
    They present subtle but fundamental attack vectors while
    also exposing significant concurrency and compatibility
    issues.
    \system appropriately handles all these cases and
    identifies several issues not mentioned by prior
    work~\cite{Connor:PKU:2020}.

    %\todo{I did not integrate the signal to work with ephemeral}
    %\input{esyscall}

    % --------------------
    \paragraph{Multithreading Design}
    %\ndd{not sure if this is here or in MT section}
    % --------------------
    So in the first version we had single threaded and the kernel delivers the
    signal to the interrupted thread. If the kernel is delivering on the backend
    of a syscall then we always come from domain 0 and thus the kernel can
    deliver to key 0 secure stack, but the problem happens if the dom 1 was
    interrupted and a signal comes. The kernel copies the pkru value and thus
    can't pus to the dom0 stack. So it faults.
    Kernel couldn't do copy to user.

    % --------------------
    To solve this we put the signal deliver into an untrusted trampoline page so
    that the kernel could always write and that would jump directly into IV to
    handle it. This worked but created the issue of a signal spoofing attack
    because now an untrusted domain could jmp to the untursted stack trampoline.
    So we solved with a nexpoline type solution..

    % --------------------
    We have a new problem with multithreading in that this open page won't work
    anymore because the page will be accessible to other threads in the same
    default domain. So we realize the interface provided by the kernel is jsut
    broken. To fix we modify the kernel to allow a return to both a registered
    stack, which is already there, and to return to a specific registered key
    value. So we now will always return to 0 domain and the 0 stack and never
    expose the data. We must then ensure that no one else registers, so we deny
    any registrations after initialization.

    To summarize: small kernel patch to allow a default domain and deny any
    registration.

  \paragraph{CET} also complicated the design of signal by adding another stack 
  that must take care of during the signal delivering. We a special syscall to 
  write to the shadow stack whcih allows us to push RIP to 
  restore address and RIP to signal handler and restore token on the shadow stack
  so we can have the required token for switching the stack when exiting \system.
  The similar trick is also used for virtualized sigreturn to switch to the old 
  stack.

  \paragraph{Multiple subdomains} As we discussed, control flow and corresponding
  CPU state are critical to the integrity of sensitive application. This applies to 
  not only the \system but also the sandbox and the safebox. Since users can run 
  whatever code in the subdomains, any interrpution during the execution of boxed code 
  can be exploited to leak data. For this reason, we block the signal from the view of 
  subdomains. The kernel can still deliver signal to \system signal entrypoint but we will
  treat it as a signal delivered in \td and pend that signal.

%% file: figs/sigentry.tex
    \begin{figure}
    \begin{lstlisting}[numbers=left]
sig_entry:
  movq $1, __flag_from_kernel(%rip)
  erim_switch
  cmpq $1, __flag_from_kernel(%rip)
  jne __sigexit
  movq $0, __flag_from_kernel(%rip)
  call _shim_sig_entry
    \end{lstlisting}
    \caption{\textbf{Signal Entrypoint}}
    \label{fig:entrypoint}
    
    \end{figure}

%% file: figs/sigstate.tex
    \begin{figure}
      \begin{tikzpicture}
        %\draw [help lines,use as bounding box] (-1,-4) grid (7,4);

        \fill[fill=purple!20] (-1.5, 2.5) arc[start angle=90,delta angle=-180,radius=2.5] -- cycle;

        \fill[fill=yellow!20] (5.5, 3.8) arc[start angle=0,delta angle=-180,radius=2.5] -- cycle;

        %\fill[fill=green!20] (7, 2) .. controls ++(205:1) and ++(-205:1) .. (4,0); %.. controls ++(-205:1) and ++(-230:1) .. (1,-1.5) .. controls (-230:1) .. (1, -4) -- (7,-4) -- cycle;
        \fill[fill=green!20] plot[smooth,tension=1] coordinates {(7, 1.5) (4,0.5) (1.5,-1.5) (1, -3.5)} -- (7,-3.5) -- cycle;

        \node[color=gray!45,scale=1.5,font=\bfseries\sffamily,rotate=90] (utt) at (-1,0) {Untrusted};

        \node[color=gray!45,scale=1.5,font=\bfseries\sffamily] (utt) at (3,3.5) {Semi-Trusted};

        \node[color=gray!45,scale=1.5,font=\bfseries\sffamily,rotate=-45] (utt) at (5,-1.5) {Trusted};

        \node[roundnodeL] (ut) at (0,0) {UT};
        \node[roundnodeL,align=center] (ssut) at (2,2.5) {Sig\\SmiUT};
        \node[roundnodeL,align=center] (sut) at (2,-2.5) {Sig\\UT};

        \node[roundnodeL,align=center] (sst) at (4,2.5) {Sig\\SmiT};
        \node[roundnodeL,align=center] (st) at (4,-2.5) {Sig\\T};

        \node[roundnodeL] (t) at (6,0) {T};

        \def\myshift#1{\raisebox{-2ex}}
        \draw[->, bend left,postaction={decorate,decoration={text along path,text align=center,text={|\sffamily\myshift|Syscall}}}] (ut) to (t);
        \def\myshift#1{\raisebox{1ex}}
        \draw[<-, bend right,postaction={decorate,decoration={text along path,text align=center,text={|\sffamily\myshift|Signal Deliver, Syscall Ret}}}] (ut) to (t);
        \draw[->, bend left,postaction={decorate,decoration={text along path,text align=center,text={|\sffamily\myshift|Recv Signal}}}] (ut) to (ssut);
        \draw[->] (ssut) -- (sut) node [midway] {Entrypoint};
        \def\myshift#1{\raisebox{-2ex}}
        \draw[<-, bend right,postaction={decorate,decoration={text along path,text align=center,text={|\sffamily\myshift|Direct Deliver}}}] (ut) to (sut);
        \def\myshift#1{\raisebox{1ex}}
        \draw[<-,bend left,postaction={decorate,decoration={text along path,text align=center,text={|\sffamily\myshift|Recv Signal}}}] (sst) to (t);
        \def\myshift#1{\raisebox{-2ex}}
        \draw[->,bend right,postaction={decorate,decoration={text along path,text align=center,text={|\sffamily\myshift|Defer Signal}}}] (st) to (t);

        \draw[->] (sst) -- (st) node [midway] {Entrypoint};

        \draw[->, bend left,postaction={decorate,decoration={text along path,text align=center,text={|\sffamily\myshift|Recv Signal in Boxing}}}] (ut.40) to (sst.250);

%        \node[fill=purple!20, 
      \end{tikzpicture}
      \caption{\textbf{State Transition with Signal;} UT:Untrusted; T: Trusted; Sig: Signal Handler, Signal masked by Kernel; Smi: Semi-Trusted Domain}
      \label{fig:state}
    \end{figure}
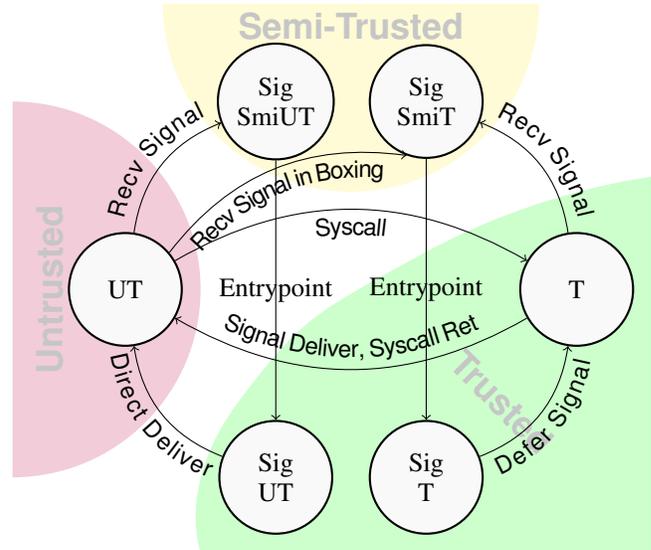